\documentclass{aa}
\usepackage{nccmath}
\usepackage{graphicx}
\usepackage{txfonts}
\usepackage{textcomp}
\usepackage{natbib}
\bibpunct{(}{)}{;}{a}{}{,}
\usepackage{amsmath}
\usepackage{color}
\usepackage{graphics}
\graphicspath{}
\usepackage{multirow}
\newcommand\phn{\phantom{0}}
\usepackage{geometry}
\usepackage{amssymb}
\usepackage{graphicx}
\usepackage{epstopdf}
\usepackage{pstricks}
\usepackage{pst-plot}
\usepackage{pstricks-add}
\usepackage{epsf}
\usepackage{pgf}
\usepackage{eqnarray,amsmath}

\begin{document} 

   \title{Deciphering the atmosphere of HAT-P-12b: solving discrepant results}

   \titlerunning{Solving contradicting results for HAT-P-12b}
   
   \author{X.~Alexoudi\inst{1}
          \and
          M.~Mallonn \inst{1}      
          \and 
          C.~von Essen \inst{2}
          \and
          J.\,D.~Turner \inst{3}
          \and
          E. Keles \inst{1}
          \and
          J.~Southworth \inst{4}
          \and
              L.~Mancini \inst{5,6,7}
              \and
              S.~Ciceri \inst{8}
              \and
                 T.~Granzer \inst{1} 
                 \and
                 C.~Denker \inst{1}
                 \and 
                 E.~Dineva \inst{1}
                 \and
         K.\,G.~Strassmeier \inst{1}  \\           
          }

   \institute{Leibniz-Institut f\"{u}r Astrophysik Potsdam (AIP), An der Sternwarte 16, D-14482 Potsdam, Germany \\
\email{xalexoudi@aip.de}
         \and
         Stellar Astrophysics Centre (SAC), Department of Physics and Astronomy, Aarhus University, Ny Munkegade 120, DK-8000, Aarhus C, Denmark
         \and
         Department of Astronomy, University of Virginia, Charlottesville, VA-22904, USA
         \and
         Astrophysics Group, Keele University, Staffordshire, ST5 5BG, UK
         \and
         Dipartimento di Fisica, Università di Roma Tor Vergata, Via della Ricerca Scientifica 1, 00133 – Roma, Italy
         \and 
         Max Planck Institute for Astronomy, Königstuhl 17, 69117 – Heidelberg, Germany
         \and
        INAF – Osservatorio Astrofisico di Torino, via Osservatorio 20, 10025 – Pino Torinese, Italy
        \and
        Department of Astronomy, Stockholm University, Alba Nova University Center, SE-106 91, Stockholm, Sweden
        }
         
   \date{Received June 21 2018 / Accepted 3 October 2018}

 
  \abstract
   {Two independent investigations of the atmosphere of the hot Jupiter HAT-P-12b by two different groups resulted in discrepant solutions. Using broad-band photometry from the ground, one study found a flat and featureless transmission spectrum that was interpreted as gray absorption by dense cloud coverage. The second study made use of Hubble Space Telescope (HST) observations and found Rayleigh scattering at optical wavelengths caused by haze.   
   }
   {The main purpose of this work is to determine the source of this inconsistency and provide feedback to prevent similar discrepancies in future analyses of other exoplanetary atmospheres. }
   {We studied the observed discrepancy via two methods. With further broad-band observations in the optical wavelength regions, we strengthened the previous measurements in precision, and with a homogeneous reanalysis of the
   published data, we were able to assess the systematic errors and the independent analyses of the two different groups.}
   {Repeating the analysis steps of both works, we found that deviating values for the orbital parameters are the reason for the aforementioned discrepancy. Our work showed a degeneracy of the planetary spectral slope with these parameters. In a homogeneous reanalysis of all data, the two literature data sets and the new observations converge to a consistent transmission spectrum, showing a low-amplitude spectral slope and a tentative detection of potassium absorption.}
   {}

   \keywords{planetary systems - planets and satellites: atmospheres - stars: individual: HAT-P-12
               }

   \maketitle
%

\section{Introduction}

In the past years, the field of exoplanetary sciences has been enriched by the numerous missions dedicated to the discovery and characterization of the planets beyond our solar system. Since the breakthrough of the first exoplanet orbiting a main-sequence star \citep{Mayor&Queloz1995}, we can count up to almost 4.000 exoplanets by mid-2018.\footnote{http://www.exoplanet.eu}

Many methods are used for the detection of planets outside our solar system. The most prevalent of them is the observation of transits. 
The term “transit” is used to define the passage of the planet in front of its host star, which for a moment blocks a portion of the transmitted starlight. 
The received radiation is diminished by an amount equal to the ratio between the disk sizes of the planet and the host star. 
Granted sufficient observational cadence and photometric precision, this method provides precise estimates of the planetary and orbital parameters (period, radius ratio, inclination, etc.).
    
Transmission spectroscopy is one of the advantages that the transit observations can offer. 
This technique is considered a fundamental tool for characterizing exoplanetary atmospheres \citep{Charbonneau2002}. 
During the transit, part of the starlight is filtered as it passes through the upper atmosphere of the planet, where the atoms and molecules can absorb or scatter accordingly, 
revealing the elemental composition of an exoplanetary atmosphere \citep{Burrows2014,Madhusudhan2016,Deming2017}. The most suitable targets for transmission spectroscopy are hot Jupiters, which orbit their host star at very close distances. The studies of this specific group of exoplanets have so far revealed a tremendous diversity of planetary atmospheres: from clear atmospheres to those that are cloudy and dominated by haze \citep[e.g.,][]{sing2016,Lendl2016,wakeford2017,Mallonnwakeford2017,Tsiaras2018,Kreidberg2018,Nikolov2018}.

The characterization of an exoplanetary atmosphere requires a very careful analysis because the signal of the planetary atmosphere is at most about 0.1\,\% of the stellar flux \citep{Deming2017}. The exoplanet community succeeded in providing robust results that were confirmed by independent teams. Examples are the clear atmosphere of WASP-39b \citep{Fischer2016,sing2016,nikolov2016}, the hazy atmosphere of GJ3470b \citep{Nascimbeni2013,Dragomir2015,Chen2017}, and the flat transmission spectrum for GJ1214b \citep{Berta2012,Kreidberg2014,Nascimbeni2015}. However, since this research takes place at the edge of what can be detected, contradicting results regarding individual planets are reported in the literature as well. Potential reasons may be the different and underestimated systematic errors, the various assumptions in the analysis (e.g., different limb-darkening laws (LDLs)), or even the time variability of the atmosphere. Examples of deviating results are the quest for potassium absorption in WASP-31b \citep{Gibson2017,Sing2015} and the debates on the spectral slope of WASP-80b \citep{Sedaghati2017,Kirk2018,Parviainen2018} and TrES-3b \citep{Parviainen2016,Mackebrandt2017}.  

\begin{figure}[h]
\includegraphics[width=9cm]{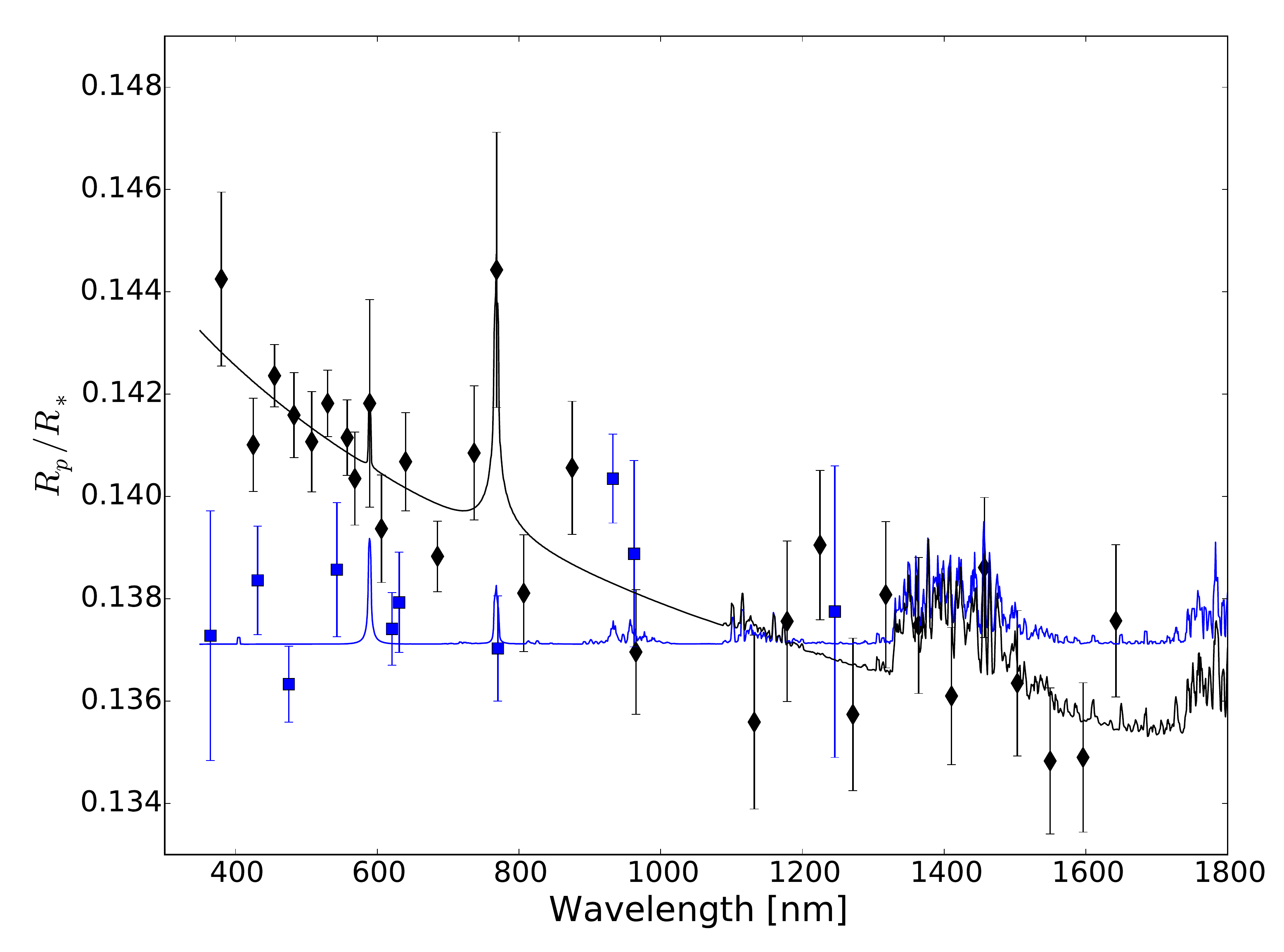}
\caption{Published transmission spectra of HAT-P-12b. The measurements of the planet-star radius ratio $R_{p}/R_{\star}$ over wavelength are presented for both works. The results of M15 are shown with blue squares, and the measurements of S16 are shown with black diamonds. Two models of \cite{fortney2010} are overplotted and present atmospheres that are dominated by clouds (blue line) and haze (black line).}
\label{plot_fig1}
\end{figure}

A specific example for a controversy in atmospheric characterization is the sub-Saturn HAT-P-12b \citep{Hartman2009}. Previous studies on the atmosphere of this exoplanet by two different groups have shown significant deviations. The first study, \cite{Mallonn2015}, hereafter M15, used ground-based broad-band photometry observations and derived a flat transmission spectrum, indicating an opaque layer of clouds as an atmospheric scenario for HAT-P-12b. The second study, carried out by \cite{sing2016}, hereafter S16, used HST spectroscopy observations and found a strong Rayleigh-scattering signature at optical wavelengths that is indicative of a haze layer.
The transmission spectra of both works are presented in Fig. \ref{plot_fig1}, with the discrepancy being evident at short wavelengths.

The goal of our work is to investigate the cause of this inconsistency regarding HAT-P-12b. The previous studies differ in several details of the analysis. Different fit routines were used, and with the aim to account for the effect of stellar limb darkening, different laws were employed by the two groups. More specifically, M15 fit the light curves with a quadratic LDL, with coefficients obtained from \cite{Claret2011}.  S16 adopted the four-parameter LDL \citep{Claret2004} and calculated new coefficients. Both studies also made different assumptions about the orbital parameters. The effect of these different assumptions on the analyses is determined and cautiously investigated in this work. We attempt to resolve the discrepancy with a homogeneous reanalysis of all the existing data sets. Moreover, in order to increase the precision at bluer wavelengths, where the discrepancy is more pronounced, we obtained additional, new transit observations.

This work is organized as follows. In Section 2 we present the new transit observations of HAT-P-12b from different facilities from the ground. In Section 3 we briefly describe the data reduction. In Section 4 we fully explain the method for the light-curve analysis and for the estimation of the error bars. In Section 5 we present the results of this work, and we identify the likely source of the inconsistency regarding HAT-P-12b as the degeneracy between the inclination and the optical spectral slope. We include a homogeneous reanalysis of the available data. In Section 6 we discuss our results and conclude with a final atmospheric characterization for this exoplanet.

\section{Data acquisition}

In order to investigate the disagreement published in the literature for the atmospheric characterization of HAT-P-12b, we collected the existing data obtained with ground-based observations  (M15) and with the HST (S16), with the aim to perform the analysis of the two different groups step by step in a homogeneous way. Additionally, we obtained data of ten new primary transit events in 2016--2017 in Johnson B and Sloan g’ filters with different facilities. Using selected broad-band filters, meter-class telescopes are able to provide sufficient accuracy to reveal general trends in a transmission spectrum, for instance, a scattering optical slope \citep{Mackebrandt2017,Mallonn2016,Dragomir2015}. 

\subsection{Published data set of M15}
The data sample of M15 included newly observed transit light curves complemented by light curves from the publications of \cite{Hartman2009}, \cite{Sada2012}, and \cite{Lee2012}. These data spanned the time range of 2007 to 2014, the wavelength range from Sloan u' band to J band in ten different filters, and were observed with ground-based 1-meter telescopes such as STELLA, medium-class telescopes such as the Telescopio Nazionale Galileo, and telescopes with mirrors of up to 8 meters such as the Large Binocular Telescope. In this work, we reanalyze the light curves from Sloan u' band to Sloan z' band and exclude partial transits. This sample of ground-based literature light curves is furthermore extended by one light curve of \cite{Turner2017} that was obtained in the Johnson B band with the 1.5-meter Kuiper-Telescope.

\subsection{Published data set of S16 }
The space-based data of HAT-P-12b, provided by S16, are a part of a comparative study of ten hot Jupiters (Program GO-12473, PI: D.K. Sing). HAT-P-12b was observed in the full optical wavelength range from 0.3 to 1.01~$\mu$m with the Space Telescope Imaging Spectrograph (STIS, \cite{hst-stis}). Additionally, for observations in the near-infrared, the Wide Field Camera 3 (WFC3, \cite{wfc3}) was also used, and the Spitzer Space Telescope Infrared Array Camera (IRAC, \cite{irac}) completed this survey with observations at 3.6 and 4.5~$\mu$m. Observations of two transits of HAT-P-12b were performed with the STIS G430L grating, covering the wavelengths from 0.29 to 0.57 $\mu$m (visit 11 and visit 12), and one observation with the G750L grating, covering the range from 0.524 to 1.027~$\mu$m (visit 22).

The team of S16 provided us with the reduced HST/STIS light curves. The visits were scheduled so that the transit is shown at the third$^\mathrm{}$ spacecraft orbit, whereas the out-of-transit baseline is contained in the$^\mathrm{}$$\text{}^\mathrm{}$ $\text{first, second, and fourth}^\mathrm{}$ orbits. Each visit consists of four telescope orbits of 96~mins. The gaps are due to the stellar target’s occultation by the Earth. A total of 24 spectroscopic light curves were acquired during the three visits (visit 11: seven light curves, visit 12: seven light curves and visit 22: ten light curves). 
The data were bias-, flat-field- and dark-corrected using the latest version of CALSTIS following a procedure described fully in \cite{Nikolov2015}.

\subsection{New observations}
We obtained new transit light curves in 2016 and 2017 from five ground-based facilities, which are the 3.5-meter Telescopio Nazionale Galileo (TNG) on the Roque de los Muchachos Observatory, the 1.23-meter and 2.2-meter telescopes of Calar Alto Observatory, the 3.5-meter Astrophysics Research Consortium (ARC) telescope from Apache Point Observatory, and the 1.2-meter STELLar Activity (STELLA) telescope \citep{Strassmeier2004}. 

\subsubsection{ARC observations}
HAT-P-12b was observed during the night of 5 July 2017 with the ARCTIC instrument \citep{arctic} on the 3.5-meter ARC telescope, Apache Point Observatory (program UV04, PI: J.D. Turner). 
ARCTIC was employed in imaging mode using the J-C B filter, and a binning of $2\times2$ pixels gave a plate scale of 0.228\arcsec/pixel. The fast mode yielded a readout time of 11s with an exposure of 45s. 

\subsubsection{TNG observations}
One transit of HAT-P-12b was observed during the night of 18 May 2017 using the DOLORES instrument \citep{dolores} on Telescopio Nazionale Galileo (TNG) with a Johnson B filter (Director's Discretionary Time (DDT) program A35DDT2, PI: X. Alexoudi). With 40\,s of exposure time, we achieved a signal-to-noise ratio (S/N) per exposure of approximately 1800. Taking into account the read-out time (25\,s) and the image transfer time (6\,s), we achieved a 71-second cadence that corresponds to a duty cycle of 56 $\%$. We defocused the telescope on purpose to about 4\arcsec full width at half-maximum (FWHM) of the object point-spread function (PSF) to avoid saturation of the detector. No windowing was applied, and the field of view (FOV) was $8.6\arcmin \times 8.6\arcmin$.

\subsubsection{CALAR ALTO 2.2m observations}
On 18 May 2017, we were also awarded with DDT (Program DDT.S17.166, PI: X. Alexoudi) to observe with the Calar Alto 2.2-meter telescope with CAFOS \citep{cafos}. 
We used the Johnson B filter with an exposure time of 60\,s. With a binning of $2\times2$ pixels and an appropriate windowing of the region of interest, 
we concluded to a cadence of approximately 80\,s and therefore a duty cycle of 73$\%$. The telescope was again defocused to about 4\arcsec FWHM to avoid saturation of the detector.

\subsubsection{Calar Alto 1.23m observations}
Two complete transit events of HAT-P-12b were observed between March and May 2017 using the Zeiss 1.23-meter telescope, which is equipped with the DLRMKIII \citep{dlr-mk3} camera ($4000 \times 4000$ pixels of 15\,$\mu$m size) and has an FOV of $21.5\arcmin \times 21.5\arcmin$. The two transits were observed using a Johnson B filter and adopting the defocussing technique for using longer exposure times (120-150\,s). This
significantly improved the quality of the photometric data. The telescope was autoguided, and the CCD was windowed to decrease the readout time.

\begin{table*}
\begin{center}
\caption{Overview of the new observations of transits of HAT-P-12b during 2016 and 2017. The columns present the observational date, the telescope, and the associated filter that was used for each observation. Furthermore, the exposure time in seconds $t_{\mathrm{exp}}$, the individual data points of each transit $N_{\mathrm{data}}$, and the scatter of the data points as root-mean-square (rms) of the light curve residuals. The $\beta$ factor is shown in the last column.}
\begin{tabular}{llccccc}
\hline
\hline
\noalign{\smallskip}
Date & Telescope &  Filter &  $t_{\mathrm{exp}}$ (s) & $N_{\mathrm{data}}$ &  rms (mag) &  $\beta$ \\
\hline
\noalign{\smallskip}
\rule[-2mm]{0mm}{4mm}
07 May 2016  &  STELLA   &g'            & \phn  80   & 163  &  0.0016    & 1.00   \\
\rule[-2mm]{0mm}{4mm}
18 Mar 2017  &  1.23m Calar Alto   &B   &   150  & \phn97   &  0.0016    & 1.22  \\
\rule[-2mm]{0mm}{4mm}
21 Mar 2017  &  STELLA   &g'            & \phn  80   & 108  &  0.0016    & 1.15  \\
\rule[-2mm]{0mm}{4mm}
02 May 2017  &  STELLA   &g'            &   \phn80   & 149  &  0.0015    & 1.34  \\
\rule[-2mm]{0mm}{4mm}
05 May 2017  &  STELLA   &g'            & \phn  80   & 130  &  0.0017    & 1.48 \\
\rule[-2mm]{0mm}{4mm}
18 May 2017  &  2.2m Calar Alto  &  B     & \phn  60   & 161  &  0.0022  & 1.15  \\
\rule[-2mm]{0mm}{4mm}
18 May 2017  &  TNG     &B              &  \phn 40   & 257  &  0.0008    & 1.15  \\
\rule[-2mm]{0mm}{4mm}
18 May 2017  &  1.23m Calar Alto  &  B    &   120  & 170  &  0.0019    & 1.30  \\
\rule[-2mm]{0mm}{4mm}
03 June 2017  &  STELLA   &g'           &  \phn 80   & 150  &  0.0017    & 1.25  \\
\rule[-2mm]{0mm}{4mm}
05 July  2017  &  ARC   &B               &  \phn 45   & 210  &  0.0017    & 1.00 \\
\hline   
\end{tabular}
\label{tab_overview}
\end{center}
\end{table*}

\subsubsection{STELLA observations}
In total, five transit observations of HAT-P-12b were obtained with STELLA with a Sloan g’ filter in 2016 and 2017. All observations were defocused to about 3\arcsec\ FWHM for the PSF. We reduced the available FOV of $22\arcmin \times 22\arcmin$ by a read-out window of about $15\arcmin \times 15\arcmin$ to shorten the readout time.

\section{Data reduction of ground-based data}
\begin{figure*}[h]
\centering
\includegraphics[width=18cm,height=10cm]{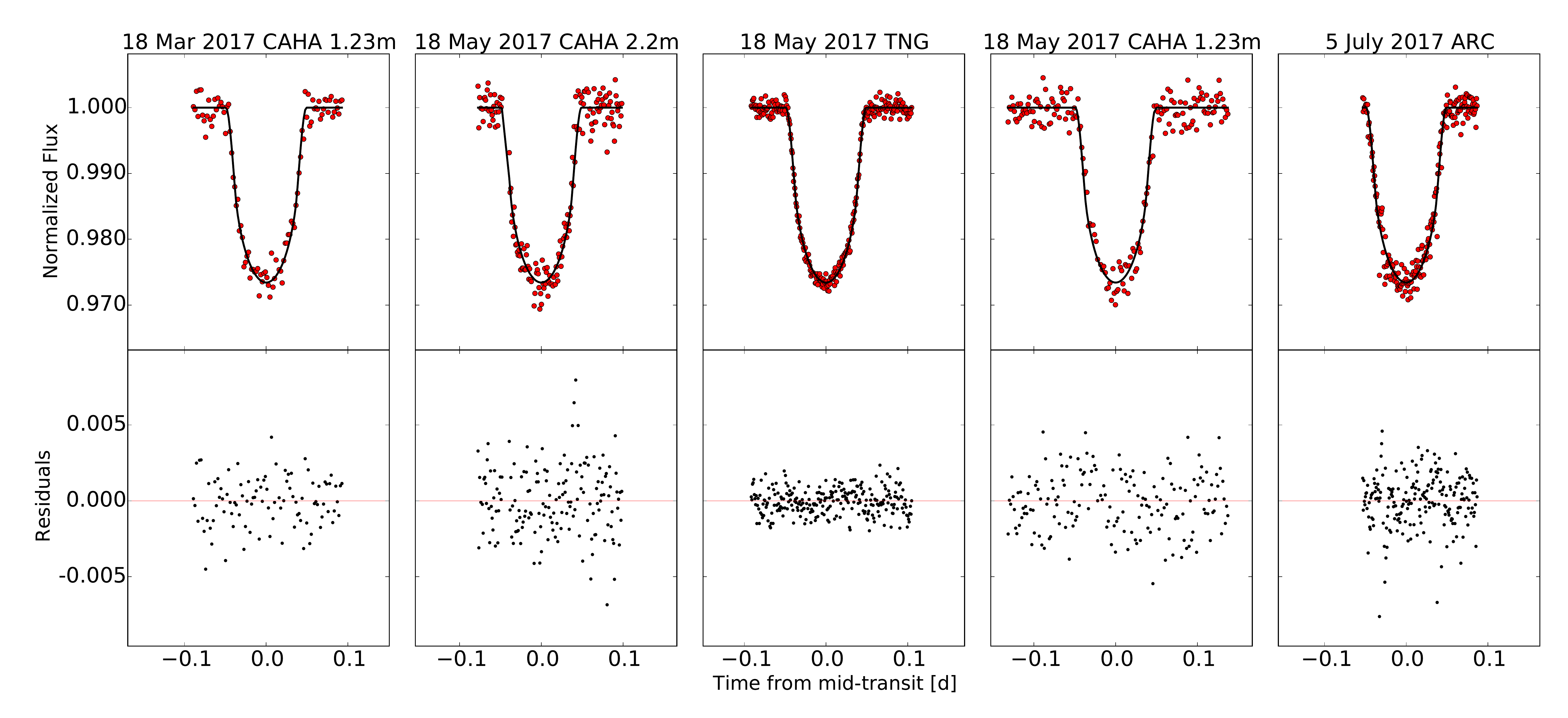}
\caption{Newly obtained B band light curves of the transit event of HAT-P-12b (upper panels). The light curves are fit and detrended according to the description of the homogeneous reanalysis in Section \ref{sec_homoreana}. Associated residuals of each light curve (lower panels).}
\label{new_lcs1}
\end{figure*}

\begin{figure*}[h]
\centering
\includegraphics[width=18cm,height=10cm]{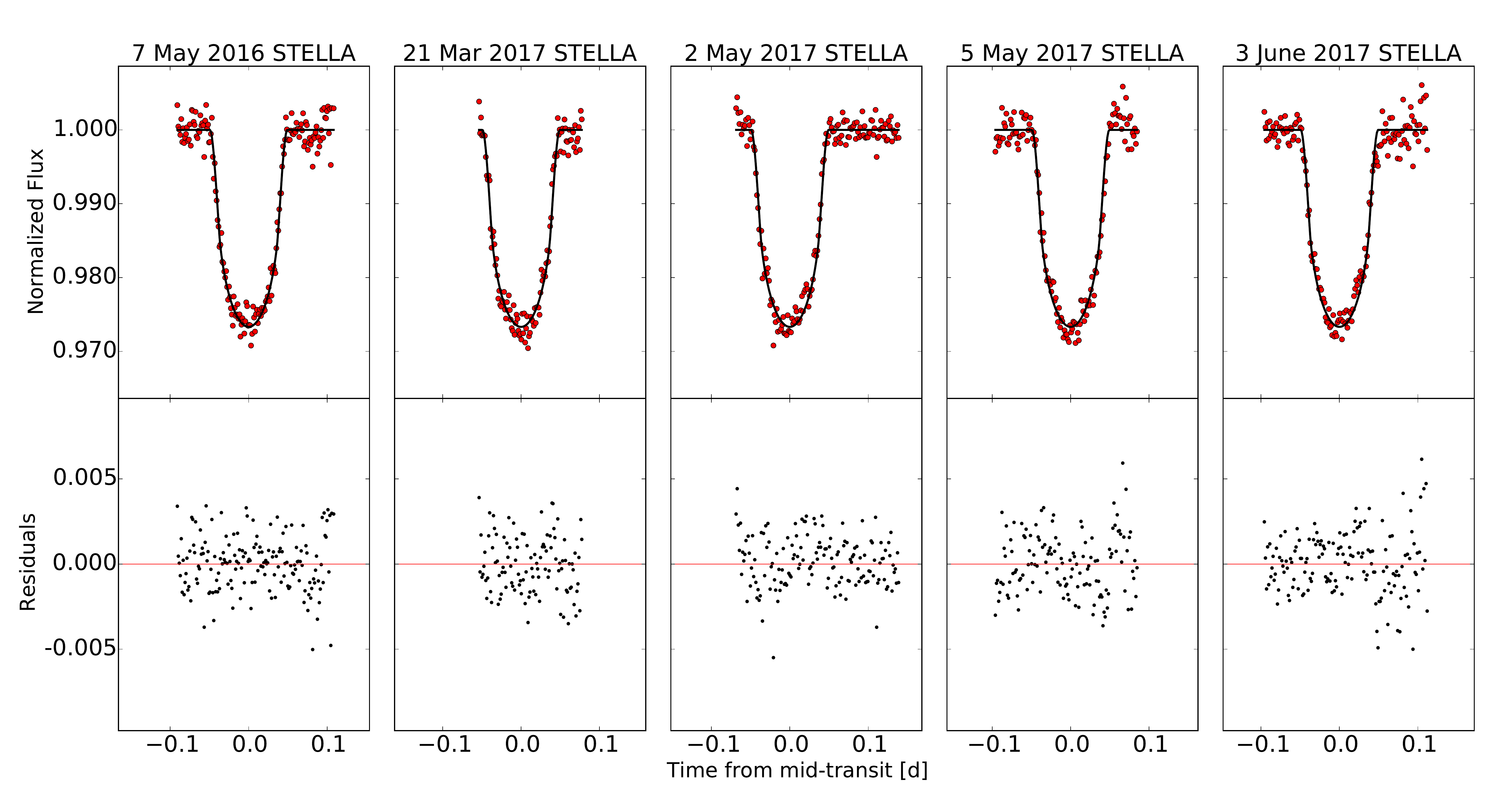}
\caption{Newly obtained g' band light curves (upper panels), fit and detrended as described in Section \ref{sec_homoreana} (similar to Fig. \ref{new_lcs1}). Their residuals are shown in the lower panels.} 

\label{new_lcs2}
\end{figure*}

The data reduction of the majority of the new observations was carried out using a customized ESO-MIDAS pipeline, which deploys the photometry software SExtractor \citep{Bertin96}. The procedure is fully described in previous works of \cite{Mallonn2015, Mallonn2016}. The only exception was the ARC observation, which was reduced using the software described in \cite{Turner2016} and \cite{Pearson2014}. 
Table \ref{tab_overview} shows the new observations as obtained from the different facilities in detail. The light curves were acquired with Johnson B and Sloan g’ filters (Figs. \ref{new_lcs1} and \ref{new_lcs2}). The rms scattering of the residuals in most cases is less than 2~mmag. The correlated noise measured with the $\beta$ factor (see Section \ref{sec_betafac}) is close to unity in the majority of our time series.

\section{Data analysis}  
\subsection{Treatment of photometric errors}
\label{sec_betafac}

A realistic assessment of the individual photometric error bars of the transit light curves is mandatory to finally derive realistic uncertainties on the transit parameters. We derived photometric uncertainties, based on the assumption of white noise, from the photometry software SExtractor \citep{Bertin96} as regular part of the data reduction. These uncertainties yield a $ \chi^2_{\mathrm{red}} $ larger than unity when an initial transit model was fit. This is an indicator that the individual photometric error bars are underestimated. Therefore, we enlarged the photometric error bars by a factor that forces $\chi^2_{\mathrm{red}}=1$. The obtained light curves often suffer from correlated (red) noise, in addition to the photon noise. The red noise is produced by a combination of systematic errors, such as changes in the atmospheric conditions, tracking of the telescope, or relative flat-field errors. Thus, adjacent points of a light curve become correlated \citep{Pont2006}. This kind of correlated noise leads to underestimated errors of the transit parameters and inconsistent parameterization of the light curves. To quantify the influence of red noise in our data, we used the concept of “$\beta$ factor” \citep{Winn08}:

\begin{ceqn}
\begin{align}
\beta =  \frac{\sigma_{\textit{r}}}{\sigma_{\textit{N}}} \,,
\label{beta}
\end{align}
\end{ceqn}
\noindent
where $\sigma_r$ is the standard deviation of the binned residuals into M bins of N points, and $\sigma_N$ is the expected standard deviation according to
\begin{ceqn}
\begin{align}
\sigma_N =  \frac{\sigma_1}{\sqrt{N}} \ \sqrt{\frac{M}{M-1}} ,
\label{sigma_n}
\end{align}
\end{ceqn}
\noindent
where $\sigma_1$ is the standard deviation of the unbinned residuals. In absence of correlated noise, $\beta$ equals unity. In the presence of correlated noise, the standard deviation of the binned data is different by a factor $\beta$ from the theoretically expected one \citep{vonEssen2013}. The $\beta$ value depends on the bin size. Therefore, we binned the residual light curves using ten binning widths of 0.3 to 1.3 times the duration of ingress. The final $\beta$ value
is the average of these ten individual values. Because of statistical fluctuations, the $\beta$ value might sometimes be lower than 1. For these cases, we set $\beta=1$ as a default. Finally, we enlarged the individual photometric error bars by the corresponding amount of $\beta$ for all ground-based new and literature light curves.

The HST/STIS light curves show too few data points per orbit to reasonably apply the method of the binned residuals, therefore we adopted $\beta\,=\,1$ for the HST data. The photometric noise was verified to yield a $\chi^2_{\mathrm{red}}=1$.

\subsection{Light-curve fitting}

\subsubsection{Transit modeling}
\label{sectransmodel}
We made use of the Python library PyAstronomy\footnote{https://github.com/sczesla/PyAstronomy}, which is a collection of packages related to astronomy developed by the PyA team at the Hamburger Sternwarte. The PyAstronomy packages contain the analytical transit model formulae of \cite{Mandel&Agol2002} and a class that allows for multiple light-curve fitting in a simultaneous way. The model of \cite{Mandel&Agol2002} is validated for light-curve modeling with the main advantage that it is capable, when used in a fitting procedure, to extract the most suitable system parameters and to investigate their possible correlations. This model is appropriate for systems with a circular or Keplerian orbit. It considers the planet and the star as spheres, while the effect of the limb darkening is also taken into account by using either the quadratic or the four-parameter LDLs. For HAT-P-12b, we adopted a circular orbit for the system, as derived from radial velocity (RV) data and reported in \cite{Hartman2009} and \cite{Knutson2014}. The parameters that constructed the model were the orbital period $P$, the semi-major axis in units of stellar radius $a/R_{\star}$, the inclination of the system $i$, the radius of the planet over the radius of the star $k = R_p/R_{\star}$, the limb-darkening coefficients, and the time of the mid-transit T0. The initial values for these parameters were adopted from the literature, from the works of \cite{Hartman2009} and M15. We used the ephemeris given in M15.
Throughout this work, we employ a Markov chain Monte Carlo (MCMC) fitting approach, enabled by PyAstronomy. Each fit consists of 300000 MCMC iterations, rejecting the first 50000 samples as burn-in phase, ensuring that the final sample is extracted from a well-converged MCMC chain, where the deviance is minimized yielding the best-fit solution. To examine the variety of probable solutions for each free-to-fit parameter, we assumed conservatively uniform prior probability distributions. In the next step, we analyzed the posterior distributions of each parameter and obtained the mean values along with their standard deviations. The errors given throughout this work correspond to 68.3\% highest probability density (HPD) intervals of the posterior probability distributions of each parameter, denoting the 1$\sigma$ level. We checked for convergence of the chains by splitting them into three equally sized sub-groups and verified that their individual mean agreed within 1$\sigma$.

\subsubsection{Detrending model for the ground-based data}

The physical properties of the observations, such as airmass, time, seeing, and $x$ and $y$ drifts of the pixel position, were used in different combinations to build up a variety of detrending models, as suggested by \cite{vonEssen2016}. They were fit simultaneously to the transit model. 
The choice of the best detrending function can be determined by the Bayesian information criterion \citep[BIC,][]{Schwarz1978}:
\begin{ceqn}
\begin{align}
 \mathrm{BIC}=\chi^2 + k \ln{N},
 \label{bic}
\end{align}
\end{ceqn}
\noindent
where $k$ is the number of model parameters, $N$ the number of data points, and $\chi^2$ is the chi-squared statistic. The value of $\chi^2$ is calculated as 
\begin{ceqn}
\begin{align}
  \chi^2=\sum\limits_{i=1}^n \left( \frac{\mathrm{Obs}_i\,-\,\mathrm{Mod}_i}{\mathrm{err}_i} \right) ^2\,,
  \label{chi_sq}
\end{align}
\end{ceqn}
\noindent
where $\mathrm{Obs}_i$ refers to the observed data points and $\mathrm{Mod}_i$ to the fit model, $\mathrm{err}_i$ are the photometric error bars, and $n$ denotes the total number of data points. By testing many combinations of detrending functions on our light curves, we concluded that a light-curve model $\mathrm{M(t)}$ including a second-order time-dependent polynomial 
\begin{ceqn}
\begin{align}
M(t)\,=\,M_0(t)\,[\mathrm{b_0} + \mathrm{b_1} t + \mathrm{b_2} t^2],  
  \label{time_poly_ground}
\end{align}
\end{ceqn}
\noindent
where $\mathrm{M_0(t)}$ is the model for a systematics-free transit model described in Section \ref{sectransmodel}, and $\mathrm{b_0},$ $\mathrm{b_1}$ and $\mathrm{b_2}$ denote the coefficients of the parabola over time, is more efficient, as it minimizes the BIC value for the vast majority of ground-based light curves. This finding is in agreement with M15 in the specific case of HAT-P-12b and with numerous studies on photometric transit light curves obtained with small-size telescopes in general \citep[e.g.,][]{Biddle2014,Maciejewski2015,Mancini2016,Mackebrandt2017,Juvan2018}. 

\subsubsection{Detrending model for the HST data}
\label{sec_detrend_HST}

The systematics model for HST data is different in comparison to the analysis of the ground-based data. We employed a model similar to the work of \cite{Huitson2013}, \cite{Nikolov2014}, and S16. In this model, we need to take into account the flux corrections for each orbit with the fitting of a fourth-order polynomial of the orbital phase $\phi$ of the HST. Higher ($\text{fifth-}^\mathrm{}$ and $\text{sixth-}^\mathrm{}$) order polynomials are in general not suggested for parameters in a detrending model because they result in larger BIC numbers \citep{Nikolov2014}. Moreover, our detrending model takes into account the shift of the detector positions $x$ and $y$. \cite{Huitson2013} reported that linearly fitting the $x$- and $y$- offsets of the spectra results in a smaller BIC and a better fit. Furthermore, our systematics model needs to account for the out-of-transit baseline with the use of an offset $\mathrm{q_0}$ and a linear term $\mathrm{q_1}$, dependent on time. In the end, the final systematic to consider is the spectral shift $sh$ in the dispersion direction. During an observation, the spectrum does not remain at the same place, and for this reason, a pixel does not receive the same wavelength for each exposure. This creates response variations from one exposure to another. Using the cross-correlation technique with the initial first exposure spectrum, S16 determined that this spectral shift is not negligible.
Therefore, the complete systematics model we used for the detrending is

\begin{figure*}[h]
\centering
\sidecaption
\includegraphics[width=12cm]{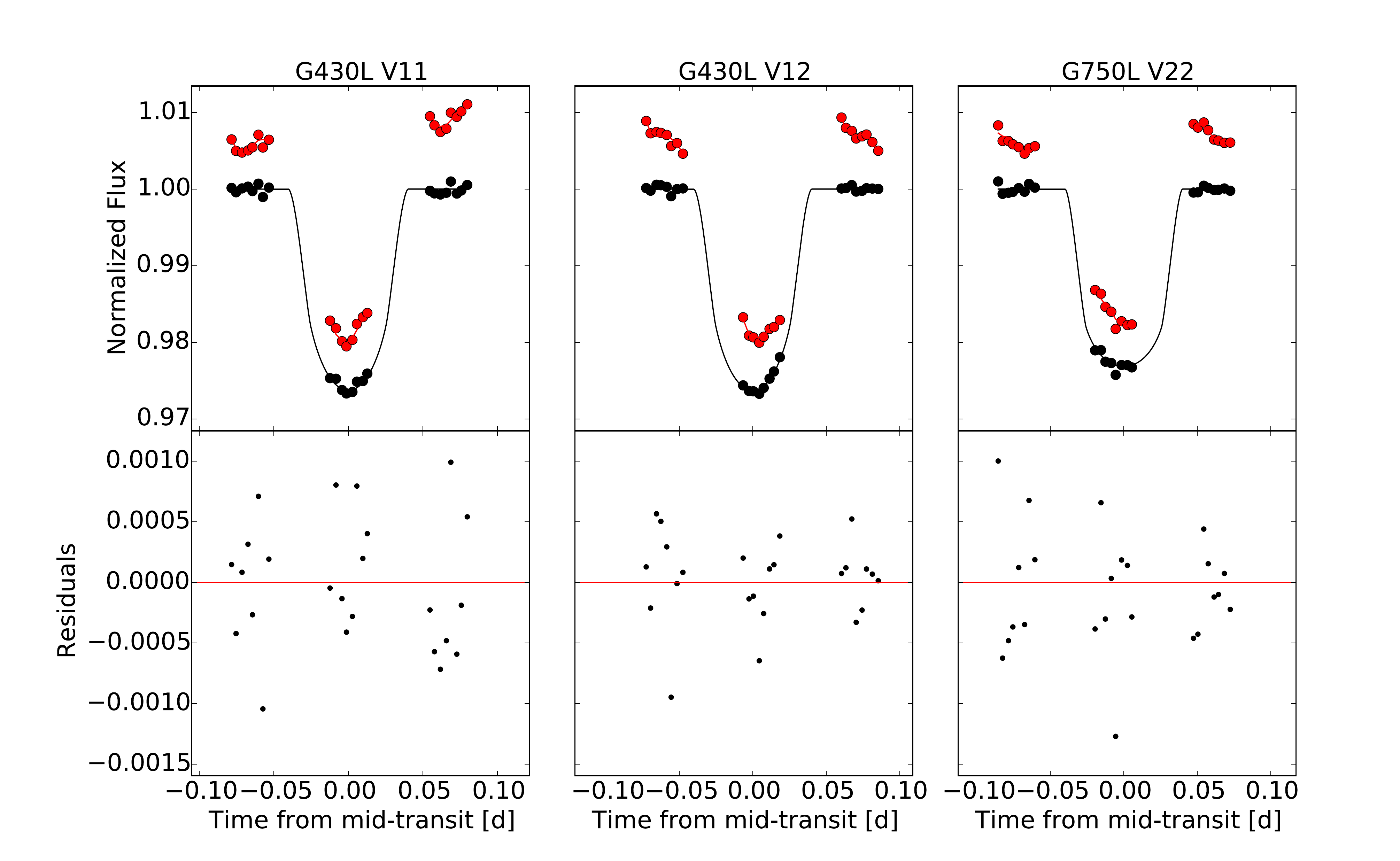}
\caption{White-light curve fitting for the HST data. The red dots show the raw data, and overplotted with the red solid line, we show the transit model according to the homogeneous reanalysis in Section \ref{sec_homoreana}. The black dots and black solid line indicate the detrended light curves. The lower panel depicts the associated residuals.}
\label{hst_whites}
\end{figure*}

\begin{ceqn}
\begin{eqnarray}
M(t)\,& = &\,M_0(t)\,[\mathrm{q_0} + \mathrm{q_1} t+ \mathrm{q_2}\phi + \mathrm{q_3}\phi^2 + \mathrm{q_4}\phi^3 + \mathrm{q_5}\phi^4 \nonumber \\
& & +\,\mathrm{q_6} sh\,+\, \mathrm{q_7} x\,+\, \mathrm{q_8} y],
    \label{time_poly_hst}
\end{eqnarray}
\end{ceqn}
\noindent
with $\mathrm{q_0 ... q_8}$, the coefficients of each term. The systematics model can be also multiplicative rather than additive \citep{Sing2018}.
As suggested by previous studies with STIS \citep{Huitson2013}, we neglected the first orbit measurements to avoid higher systematic errors. These systematics are usually caused by the telescope heating and cooling per orbit, a term known as “thermal breathing”. The temperature changes because of the expansion or contraction of the telescope, thus these are interpreted as variations in the central position of the spectrum and in the PSF. Because of these problems with the first exposure of each orbit \citep{Sing2011b}, we neglected these individual data as well. Figure \ref{hst_whites} shows the white-light curves of the three visits overplotted with our transit and detrending model.

\section{Results}
\subsection{Reanalysis of the ground-based data}

Our ground-based data set consisted of photometric transit light curves from M15, \cite{Hartman2009}, \cite{Sada2012}, \cite{Lee2012}, 1 light curve from \cite{Turner2017}, and our 10 new light curves of Table~\ref{tab_overview}. We made use of a total number of 35 light curves that were observed during a time period covering ten years (2007--2017).

\begin{figure*}[h]
\centering
\sidecaption
\includegraphics[width=12cm]{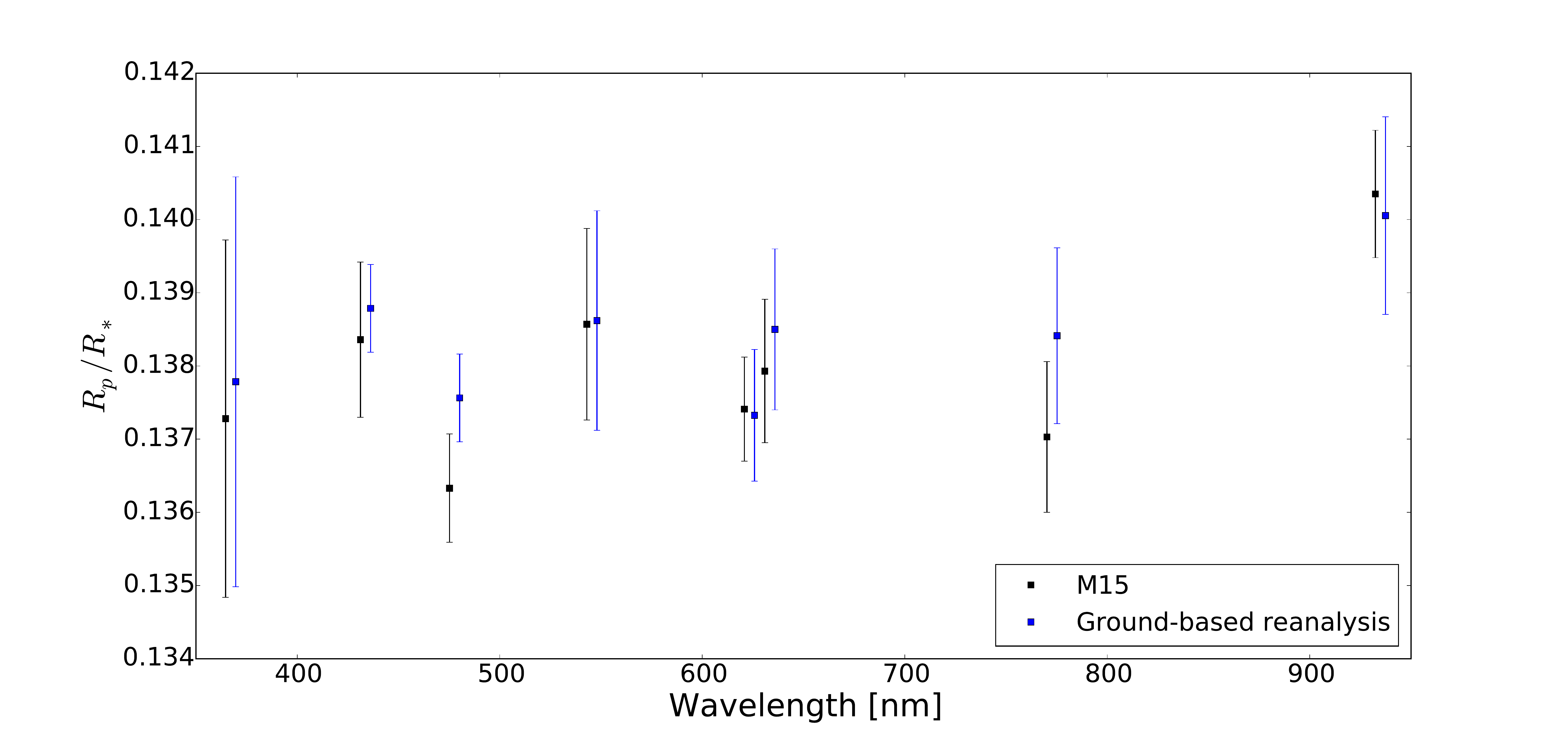}
\caption{Reanalysis of the ground-based data of HAT-P-12b and comparison with the previous results from M15. The values of this work are offset in wavelength for better visibility. We included additional data for the B and g' band data at 435~nm and 475~nm, which results in higher precision. }
\label{mallonn_comp}
\end{figure*}

As the first step of our analysis, we attempted to reproduce the broad-band filter transmission spectrum of M15. For this purpose, we adapted the details of the analysis of this work, which employs the quadratic LDL with the two coefficients $u$ and $v$. The latter was fixed to the theoretical values provided in M15, while the former was left free to fit. The transit parameters $T0$, $a/R_{\star}$, $i$, and $P$ were fixed to the values of M15. The model fit consisted of the following free-to-fit parameters: one value per broad-band filter of $k$ and $u$, along with the three constants $\mathrm{b_0, b_1,and \text{ }    b_2}$ of the time-dependent polynomial (Eq. \ref{time_poly_ground}) per individual light curve. The reanalyzed detrended light curves are presented in Figs.~\ref{new_lcs1}, \ref{new_lcs2}, \ref{u_band_final}, \ref{b_band_final}, \ref{g_band_final}, \ref{V_band_final}, \ref{r_band_final}, \ref{R_band_final}, and \ref{iz_band_final}.
From this global fit and from investigating the credibility intervals of the parameters, we find compatible results with M15 for $k$ over wavelength, as shown in Fig.~\ref{mallonn_comp}. 

To conclude this part of our work, using the same LDL and applying the same treatment of the fit parameters, using the same detrending model and the same enlargement procedure on the photometric errors, but another fitting routine, we were able to reproduce the results obtained by M15. It is important to emphasize that our work includes additional light curves from new observations in Johnson B and Sloan g’ filters, obtained with different facilities from the ground, which improved our precision in both bands. The broad-band results agree with a flat transmission spectrum of HAT-P-12b within one atmospheric pressure scale height.  
However, the discrepancy concerning the atmosphere of HAT-P-12b is not yet understood. Therefore, we will continue this investigation by taking into account the different LDLs deployed by the different groups and the different system parameters used for the light-curve analyses. 

\subsubsection{Importance of the LDLs}
The studies of M15 and S16 differ in the choice of the LDL. While the former uses the quadratic law, the latter employs the four-parameter LDL with its coefficients $\alpha1$, $\alpha2$, $\alpha3$, and $\alpha4$. To investigate the importance of this choice in the case of HAT-P-12b, we employed both the quadratic and the four-parameter LDL by fitting the ground-based data again. We used the coefficients provided by \cite{Claret2011} from stellar ATLAS models. We also investigated a plausible change in $k$ when fixing all coefficients to theoretical values in comparison to leaving the first coefficient of either the quadratic or of the four-parameter LDL free to fit. 

For all cases, the difference in $k$ is not significant. The values per individual filter change only by a fraction of our 1\,$\sigma$ uncertainties, whereas the overall slope remains unaffected.

Moreover, to investigate the suggestion by \cite{Csizmadia2013} that different theoretical calculations of the limb-darkening coefficients by individual authors might lead to varying results for $k$, we fit our light curves with limb-darkening coefficients obtained from both S16 and \cite{Claret2011}. However, the effect of the coefficients from different authors is not of significance either. The obtained \textit{k} per filter is in agreement with the published ground-based data of HAT-P-12b and supports a flat-spectrum scenario at optical wavelengths. 

The main outcome of this part of the analysis is that different LDLs and coefficients are certainly not the cause of the discrepancy in the published transmission spectra of HAT-P-12b. In a further step toward the clarification of the problem, we investigated the effect of the orbital parameters on the light-curve fitting procedure, and specifically, the influence of the system inclination.

\subsubsection{Importance of the system inclination}

\begin{figure*}[h]
\centering
\includegraphics[height=9cm]{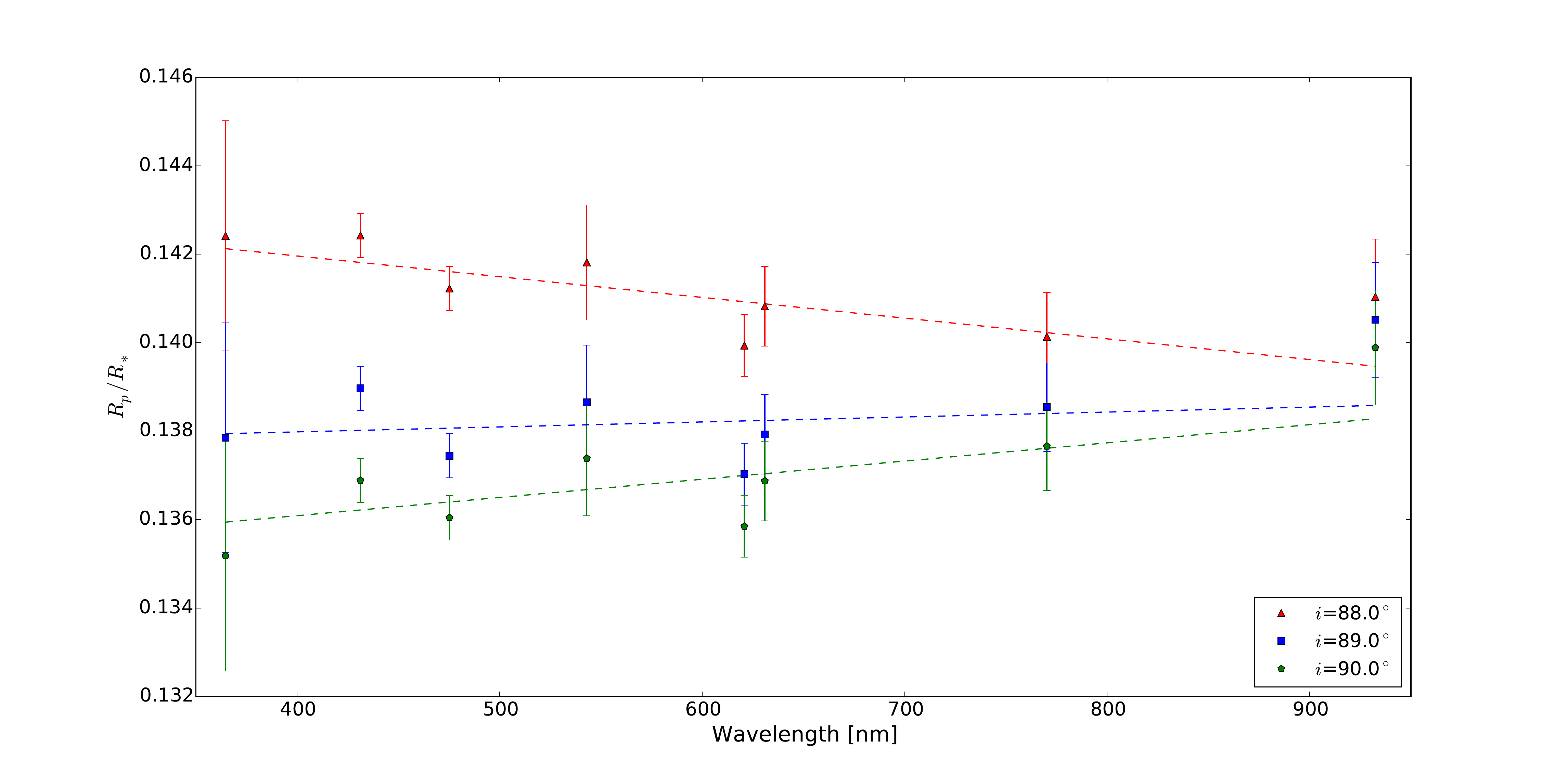}
\caption{Variation in transmission spectrum for three different inclination values. 
Different values of $i$ cause an offset in $k$, which is wavelength dependent. This effect can mimic a slope in the transmission spectrum. The dashed lines mark linear regressions.}
\label{inclinations}
\end{figure*}

Both studies, M15 and S16, fixed the orbital parameters $i$ and $a/R_{\star}$ when they fit for $k$ over wavelength. The one work fixed these parameters to values that they derived from a joint fit to all their data, the other work fixed these parameters to values derived from the HAT-P-12b Spitzer light curves of S16.
In this section, we investigate the effect of these parameters on the value of $k$ using the example of $i$.  The literature agrees (Table \ref{transitparam_H12}) on an inclination value for the HAT-P-12 system of about 89° \citep{Hartman2009,Line2013,Mallonn2015,Mancini2018}. We defined three cases in total with fixed different inclinations of 88°, 89° , and 90°. We chose the four-parameter LDL with fixed coefficients \citep{Claret2011} and fit the ground-based light curves again with the aim to investigate if there is a significant variation in $k$ with respect to $i$.
 Figure \ref{inclinations} shows that changes in the fixed inclination value cause an offset in $k$, which is wavelength dependent. Therefore, a change in $i$ can cause a slope in the transmission spectrum. Potentially, the assumption of an inaccurately small inclination for HAT-P-12b could imply a Rayleigh slope. 

This dependency of the optical slope on the orbital parameters is briefly described in \cite{Pont2013}. With our investigation, we verify that the discrepancy between M15 and S16 is potentially based on this degeneracy of $i$ and $k$. By a joint fit of all ground-based light curves with free transit parameters $i$, $a/R_{\star}$, and $k$, we derive a refined value for the inclination, $i\,=\,88.83^\circ\,\pm\,0.19^\circ$. 

\subsection{Reanalysis of the HST data}
\label{sec_HST_reana}
In this part of our work, we test if the sloped HST/STIS transmission spectrum can be reproduced, following the analysis procedure described in S16 step by step. Because the HST light curves lack observational data during transit ingress and egress, they are not suited to accurately constrain the orbital parameters $i$ and $a/R_{\star}$. Therefore, S16 derived these parameters from the two Spitzer light curves at 3.5 and 4.6~$\mu\mathrm{m}$ included in their work, resulting in $i_{\mathrm{S16}}\,=\,88.10^\circ\,\pm\,0.27$ and $a/R_{\star,\mathrm{S16}}\,=\,11.19\,\pm\,0.22$ (Nikolov, private communication). 

We derived the STIS transmission spectrum from the set of light curves over wavelengths as supplied to us by the team of S16, keeping $i$ and $a/R_{\star}$ fixed to the Spitzer values. A detrending model similar to S16 was employed (see Section \ref{sec_detrend_HST}), and as carried out by S16, we used the four-parameter LDL with coefficients fixed to theoretical values. The free-to-fit parameters were $k$ per wavelength channel and detrending coefficients per light curve. The detrended light curves are presented in Figs. \ref{hst_i88a} and \ref{hst_i88c}.  

\begin{figure*}[h]
\centering
\includegraphics[height=9cm]{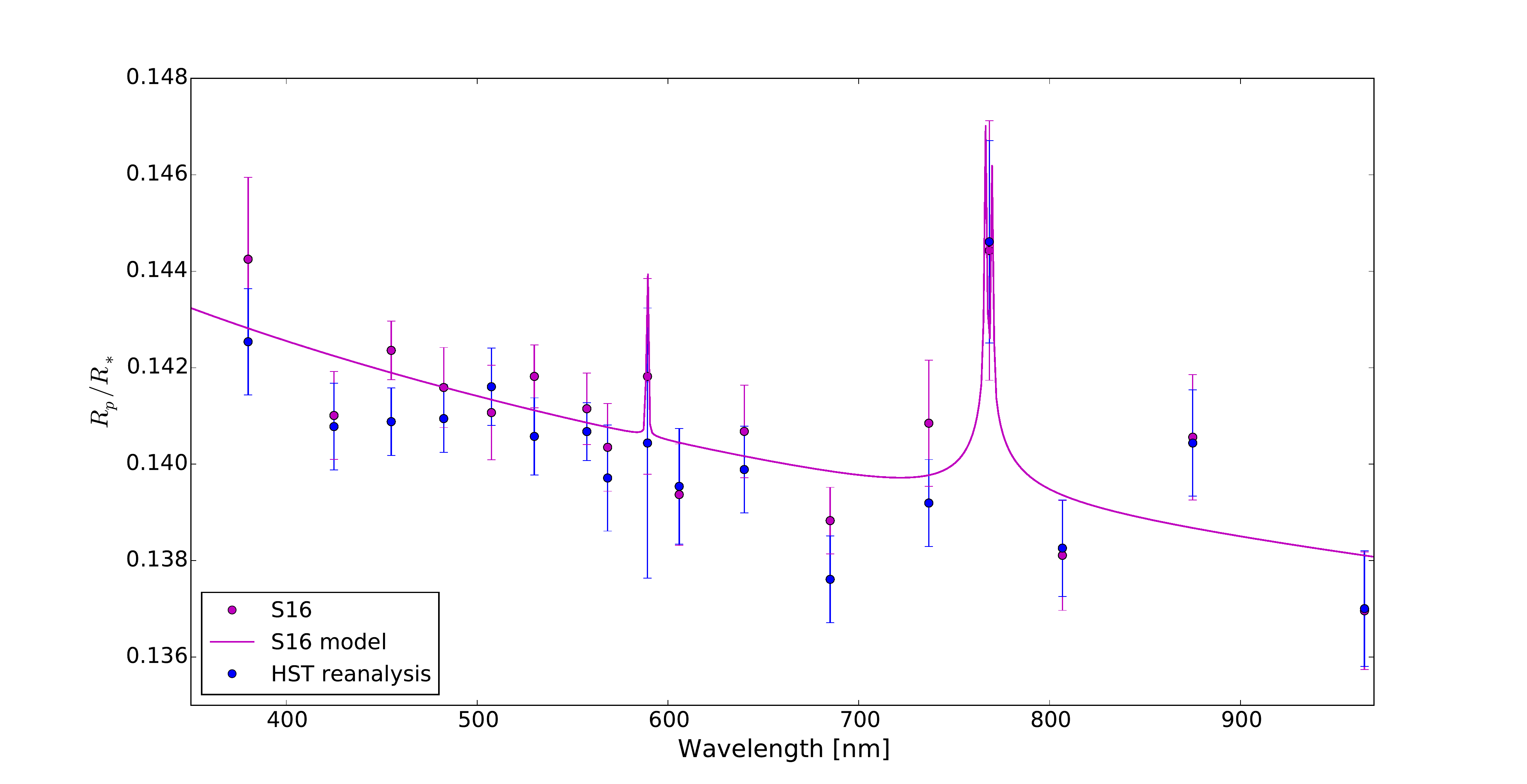}
\caption{Reanalysis of the HST data using the orbital parameters derived by S16 using the Spitzer transit light-curves. The reproduced transmission spectrum (blue data points) agrees well with the transmission spectrum of S16 (magenta points).}
\label{hst_reanalysis}
\end{figure*}

Figure \ref{hst_reanalysis} shows the resulting reanalyzed transmission spectrum, which agrees very well with the spectrum presented in S16. We summarize that S16 suggested slightly different orbital parameters than were given by a fit to the numerous ground-based data and in the literature. If these parameter values are used to derive the transmission spectrum, we reproduce an evident slope of $k$ over wavelength.

\begin{table*}
\caption{Transit parameters of HAT-P-12b derived in this work and comparison to previously published values.}
\label{transitparam_H12}
\begin{center}
\begin{tabular}{lccccr}
\hline
\noalign{\smallskip}
References    &  $a/R_{\star}$ &  $k\,=\,R_p/R_{\star}$ &  i (deg) \\
\hline
\noalign{\smallskip}
\cite{Hartman2009} &  11.76 $\pm$ 0.18    &  0.1406 $\pm$ 0.0013  & 89.0 $\pm$ 0.4 & & \\
\cite{Lee2012}     &  11.74 $\pm$ 0.17    &  0.1370 $\pm$ 0.0019  & 89.9 $\pm$ 0.1  & &   \\
\cite{Sada2012}    & 11.22 $\pm$ 0.57    &  0.1404 $\pm$ 0.0026  & 88.5 $\pm$ 1.0  & &  \\
\cite{Line2013}    & 11.60 $\pm$ 0.39    &  0.1370 $\pm$ 0.0011   &  88.7 $\pm$ 0.6  & &  \\
\cite{Mallonn2015}  & 11.72 $\pm$ 0.12 & 0.13779 $\pm$ 0.00079 & 88.98 $\pm$ 0.29 & & \\
\cite{sing2016}  & 11.19 $\pm$ 0.22 & 0.13596 $\pm$ 0.00059 & 88.10 $\pm$ 0.27 & & \\
\cite{Mancini2018} &11.93  $\pm$ 0.28  &0.13898 $\pm$0.00069 & 89.10 $\pm$ 0.24\\
\hline
\textbf{This work}  & 11.68 $\pm$ 0.12 & 0.13847$\pm$ 0.00054 & 88.83 $\pm$ 0.19 & & \\
\hline
\end{tabular}
\end{center}
\end{table*}

\subsection{Homogeneous reanalysis of all data}
\label{sec_homoreana}
In the previous parts of this work, we successfully reproduced the published results from M15 and S16 with the analyses of each data set independently and according to the published procedure of each group. Now, we aim for a homogeneous reanalysis of both the ground-based and the space-based data. 
We consider the orbital parameter values derived in this work (Table \ref{transitparam_H12}) as the most accurate ones because they are drawn from a very large sample of independently observed data. The remaining systematic noise inherent to the individual light curves of this sample is expected to average out. Moreover, these values broadly agree with the literature values published previously, while the S16 values deviate by about 2\,$\sigma$. The source of this slight deviation might be pure statistical fluctuation and is not investigated further here. Thus, we fixed $i$ and $a/R_{\star}$ to the best-fit ground-based values, and let $k$ vary for in total 17 wavelength channels of the HST data and eight broad-band filters of the ground-based data. Each individual light curve has its own independent and simultaneously fit detrending function. The wavelength-dependent limb-darkening coefficients of the four-parameter LDL were fixed to the theoretical values from \cite{Claret2011}. The result, that is, the final transmission spectrum, is presented in Fig. \ref{final_spectr} and Table \ref{final_analys}. The $k$ values of ground- and space-based data finally agree broadly.

\begin{table*}
\caption{Final results of the homogeneous reanalysis described in Section \ref{sec_homoreana} with the corresponding limb-darkening coefficients.}
\label{final_analys}
\begin{center}
\begin{tabular}{p{25mm}p{30mm}p{15mm}p{15mm}p{15mm}p{15mm}}
\hline
\hline
\noalign{\smallskip}
$\lambda$ range (nm)   &$k\,=\,R_p/R_{\star}$ (HST)  &{$\alpha1$} &{$\alpha2$} &{$\alpha3$} &{$\alpha4$} \\
\hline
\noalign{\smallskip}
 350.0 - 410.0 & 0.14022 $\pm$ 0.00110  & 0.7688 & -- 1.2583 & 1.8781 & -- 0.4346 \\
 410.0 - 440.0 & 0.13836 $\pm$ 0.00088  & 0.6393 & -- 0.9882 & 1.9149 & -- 0.6358 \\
 440.0 - 470.0 & 0.13877 $\pm$ 0.00067  & 0.5994 & -- 0.8784 & 1.8883 & -- 0.6953 \\
 470.0 - 495.0 & 0.13855 $\pm$ 0.00071  & 0.5903 & -- 0.8136 & 1.8253 & -- 0.7031 \\
 495.0 - 520.0 & 0.13808 $\pm$ 0.00077  & 0.5958 & -- 0.7606 & 1.7314 & -- 0.6821 \\
 520.0 - 540.0 & 0.13904 $\pm$ 0.00077  & 0.6008 & -- 0.7129 & 1.6469 & -- 0.6631 \\
 540.0 - 575.0 & 0.13874 $\pm$ 0.00062  & 0.6052 & -- 0.6577 & 1.5473 & -- 0.6385 \\
 550.0 - 586.8 & 0.13801 $\pm$ 0.00108  & 0.6058 & -- 0.6380 & 1.5105 & -- 0.6276 \\
 586.8 - 591.8 & 0.13846 $\pm$ 0.00268  & 0.6070 & -- 0.6002 & 1.4400 & -- 0.6068 \\
 591.8 - 620.0 & 0.13713 $\pm$ 0.00115  & 0.6079 & -- 0.5701 & 1.3839 & -- 0.5903 \\
 620.0 - 660.0 & 0.13798 $\pm$ 0.00094  & 0.6311 & -- 0.5640 & 1.2853 & -- 0.5517 \\
 660.0 - 710.0 & 0.13660 $\pm$ 0.00085  & 0.6472 & -- 0.5788 & 1.2450 & -- 0.5319 \\
 710.0 - 763.4 & 0.13700 $\pm$ 0.00105  & 0.6657 & -- 0.5959 & 1.1987 & -- 0.5092 \\
 763.4 - 773.4 & 0.14244 $\pm$ 0.00255  & 0.6670 & -- 0.6063 & 1.1704 & -- 0.4953 \\
 773.4 - 840.0 & 0.13622 $\pm$ 0.00098  & 0.6789 & -- 0.6034 & 1.1400 & -- 0.4821 \\
 840.0 - 910.0 & 0.13867 $\pm$ 0.00111  & 0.6815 & -- 0.5972 & 1.0860 & -- 0.4587 \\
 910.0 - 1020.0 & 0.13516 $\pm$ 0.00121 & 0.6848 & -- 0.5889 & 1.0150 & -- 0.4278 \\
\hline
Filter &$k\,=\,R_p/R_{\star}$ (Ground) &{$\alpha1$} &{$\alpha2$} &{$\alpha3$} &{$\alpha4$} \\
\hline
u' & 0.13803 $\pm$ 0.00264 & 0.8321 & -- 1.3903 & 1.8601 & -- 0.3362 \\
B  & 0.13914 $\pm$ 0.00055 & 0.6097 & -- 0.9264 & 1.9233 & -- 0.6819 \\
g' & 0.13765 $\pm$ 0.00050 & 0.5887 & -- 0.8286 & 1.8520 & -- 0.7091 \\
V  & 0.13881 $\pm$ 0.00136 & 0.6047 & -- 0.6754 & 1.5804 & -- 0.6482 \\
r' & 0.13779 $\pm$ 0.00076 & 0.6087 & -- 0.5439 & 1.3350 & -- 0.5759 \\
R  & 0.13816 $\pm$ 0.00095 & 0.6293 & -- 0.5623 & 1.2899 & -- 0.5539 \\
i' & 0.13872 $\pm$ 0.00116 & 0.6775 & -- 0.6068 & 1.1691 & -- 0.4947 \\
z' & 0.14071 $\pm$ 0.00135 & 0.6849 & -- 0.5888 & 1.0138 & -- 0.4273 \\
\hline
\hline
\end{tabular}
\end{center}
\end{table*}

\subsection{Physical parameters and the ephemeris of HAT-P-12b}
\label{sec_physpara}
Based on our newly derived transit parameters, we refined the planetary properties using the equations from \cite{Southworth2007}, \cite{Southworth2010}, \cite{Winn10}, \cite{Seager2011}, and \cite{Turner2016}:

\begin{ceqn}
\begin{align}
M_p = \left(\frac{\sqrt{1-e^2}}{28.4329}\right) \left(\frac{K_{\star}}{\mathrm{sin}\,i} \right)\left(\frac{\textit{P}}{1\,\mathrm{yr}}\right)^{1/3} \left(\frac{M_{\star}}{M_\odot} \right)^{2/3} M_{\mathrm{Jup}},
\label{equ_Mp}
\end{align}
\end{ceqn}
\noindent
where $K_{\star}=35.8 \pm 1.9$~km\,s$^{-1}$ is the RV semi-amplitude of the star and $M_{\star} = 0.73$~M$_\odot$ \citep{Hartman2009}. M$_\odot$ denotes one solar mass. The variable $e$ stands for the eccentricity of the system, which in case of HAT-P-12b is assumed to be zero.
The surface gravitational acceleration is 

\begin{ceqn}
\begin{align}
g_p=\frac{2\pi}{\textit{P}} \left( \frac{a}{R_p} \right)^2 \frac{\sqrt{1-e^2}}{\mathrm{sin}\,i} K_{\star},
  \label{equ_gp}
\end{align}
\end{ceqn}
\noindent
and the modified equilibrium temperature $T_{\mathrm{eq}}$ is given by
\begin{ceqn}
\begin{align}
T_{\mathrm{eq}}=T_{\mathrm{eff}} \left( \frac{R_{\star}}{2\,a} \right)^{1/2},
  \label{equ_Teq}
\end{align}
\end{ceqn}
\noindent
where $T_{\mathrm{eff}}$ is the effective temperature of the host star. The formula is derived under the assumption that $A=1-4F,$ with $A$ the Bond albedo and $F$ the heat redistribution factor \citep{South2010}.

The ability of the planet to interact with other bodies in its environment is measured by the Safronov number $\Theta$ \citep{Safronov1972}: 

\begin{ceqn}
\begin{align}
\Theta = \frac{M_p\,a}{M_{\star}\,R_p}.
  \label{safronov}
\end{align}
\end{ceqn}
\noindent
To refine the orbital ephemeris, we derived the individual mid-transit times of the new transit observations by a separated fit individually to every light curve, summarized in Table \ref{tab_ephemeris}, and combined them with the published mid-transit times from \cite{Mallonn2015}. A weighted linear least-squares fit to all the mid-transit times versus their cycle number resulted in an updated ephemeris:
\begin{equation}\label{ephemeris}
T_c = \mathrm{BJD(TDB)}\,2456446.635829(36) +  3.21305766 (13)\,N \,.
\end{equation}
\noindent
Here, $T_c$ denotes the predicted transit central time and $N$ the corresponding cycle number of the reference mid-time. The orbital period agrees with that published by \cite{Mallonn2015}, with slightly improved precision, and it shows a difference of about 1 $\sigma$ to the work of \cite{Hartman2009}.

In Table \ref{comparison_mancini} we compare our values with the most recent literature values of HAT-P-12b \citep{Mancini2018}, and our results are in agreement. Moreover, the equilibrium temperature estimated here agrees with the upper limit on the brightness temperature derived by \cite{Todorov2013} if for simplicity we assume the planet to radiate as a blackbody.

\begin{table*}
\begin{center}
\caption{Individual transit mid-times of the new transit observations of this work.}
\begin{tabular}{llcc}
\hline
\hline
\noalign{\smallskip}
Date of Observation &BJD{\textunderscore}TDB (2457000 +)\\
\hline
\noalign{\smallskip}
\rule[-2mm]{0mm}{4mm}
07 May 2016   &516.58394 $\pm$ 0.00031 \\
\rule[-2mm]{0mm}{4mm}
18 Mar 2017   &831.46339  $\pm$ 0.00037 \\
\rule[-2mm]{0mm}{4mm}
21 Mar 2017   &834.67573 $\pm$ 0.00042\\
\rule[-2mm]{0mm}{4mm}
02 May 2017   &876.44558 $\pm$ 0.00036 \\
\rule[-2mm]{0mm}{4mm}
05 May 2017   &879.66160 $\pm$ 0.00036 \\
\rule[-2mm]{0mm}{4mm}
18 May 2017   &892.51113 $\pm$ 0.00046 \\
\rule[-2mm]{0mm}{4mm}
18 May 2017  &892.51191 $\pm$ 0.00012\\
\rule[-2mm]{0mm}{4mm}
18 May 2017  &892.51258 $\pm$ 0.00041\\
\rule[-2mm]{0mm}{4mm}
03 June 2017 &908.57706 $\pm$ 0.00040\\
\rule[-2mm]{0mm}{4mm}
05 July  2017 &940.70759 $\pm$ 0.00036\\
\hline   
\end{tabular}
\label{tab_ephemeris}
\end{center}
\end{table*}
\begin{table*}
\begin{center}
\caption{Estimates of the physical properties of HAT-P-12b as derived from this work in comparison to literature values recently provided by \cite{Mancini2018}.}
\begin{tabular}{p{20mm}p{30mm}p{30mm}l}
\hline
\noalign{\smallskip}
\hline
Parameters    &This work &\cite{Mancini2018}  \\
\hline
\noalign{\smallskip}
Mass ($M_J$)         & 0.211 $\pm$ 0.012 & 0.201 $\pm$ 0.011 \\
log\,\textit{g} (cgs)  &  2.78 $\pm$ 0.03   & 2.77 $\pm$ 0.03 \\
T$_{\mathrm{eq}}$ (K)              &  957 $\pm$ 15   &  955 $\pm$ 11  \\
$\Theta$   & 0.0237 $\pm$ 0.0009& 0.0238 $\pm$ 0.0012 \\
H (km)               &  577.0 $\pm$ 12.7    & 586.0 $\pm$ 9.4 \\
\hline   
\end{tabular}
\label{comparison_mancini}
\end{center}
\end{table*}

\section{Discussion and conclusion}
\begin{figure*}[h]
\centering
\includegraphics[height=11cm]{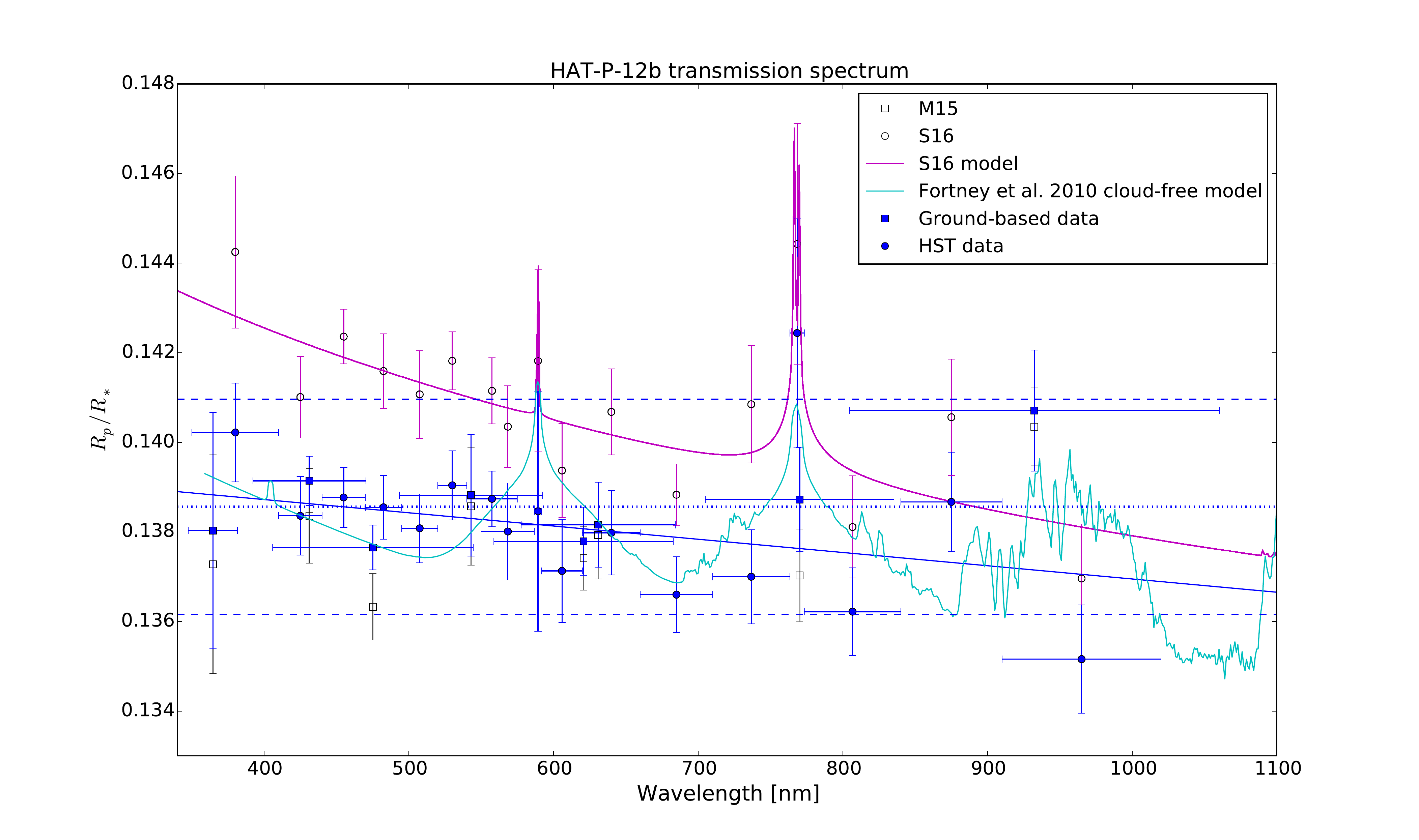}
\caption{Transmission spectrum of HAT-P-12b as derived from the homogeneous reanalysis of all data from the ground and from the HST with their associated error bars. For reference and comparison, we overplot the values obtained by M15 (black empty squares). The blue dashed lines show plus-minus two scale heights from the weighted average value of $k$ (blue dotted line). In magenta we show the values of S16 together with the suggested atmospheric model (magenta solid line). The cyan solid overplotted line represents the cloud-free, solar-composition model of HAT-P-12b from \cite{fortney2010} for comparison. The blue solid line is a linear regression of the weighted $k$ values.} 
\label{final_spectr}
\end{figure*}

The motivation for this work was to draw a homogeneous picture of the atmosphere of HAT-P-12b and to solve the discrepancy between the transmission spectra published by M15 and S16. During the analysis, we found that this discrepancy was not caused by the different analysis tools, as M15 used Levenberg-Marquardt fitting with the JKTEBOP software package \citep{Southworth08} and S16 used the MCMC fitting of the \cite{Mandel&Agol2002} transit models. Neither was the usage of different LDLs the reason, nor could it be found in subtle differences in the treatment of the limb-darkening coefficients. Following the procedures of both studies for an individual reproduction of their results, we noted a difference in the orbital planetary parameters that is the likely cause of the discrepancy. While both teams assumed the parameters $i$ and $a/R_{\star}$ to be wavelength independent and kept them fixed during their final light-curve fit, they derived different values for them. The value of M15 agreed with previous literature values, while the light-curve fit of two Spitzer light curves of S16 yielded slightly deviating orbital parameters (Nikolov, private communication). With these two sets of parameters, we succeeded in reproducing the deviating transmission spectrum. When we reanalyzed all ground-based and space-based data using newly derived values for the orbital parameters, we found consistent results among all observations. New transit light curves taken at short wavelengths increased the precision further.

HAT-P-12b is one of the ten targets of the comparative study of S16. The orbital parameters of the other targets, derived from the fit to HST white-light curves \citep{Huitson2013,Sing2013,Nikolov2014,Sing2015,Nikolov2015}, are generally in good agreement with previously published values. Therefore, HAT-P-12b probably is a particular case in the HST survey planet sample, and we currently find no indications that would make a reanalysis of other data of this sample necessary.  

To our knowledge, a degeneracy between the orbital parameters $i$ and $a/R_{\star}$ and the derived slope in the exoplanet transmission spectrum is not thoroughly discussed in the literature. Such an exercise was not foreseen for this paper either, but an extensive work based on simulated data is currently prepared by our team. Many hot Jupiters that have been investigated with transmission spectroscopy show a (negative) slope in their spectrum, which is generally interpreted as a signature of scattering by aerosol (haze) particles \citep[e.g.,][]{Jordan2013,sing2016,Mallonnwakeford2017,Chen2017}. It is extensively discussed in literature that such a slope can also be mimicked by third light of another stellar body falling to the photometric aperture and brightness inhomogeneities such as spots and faculae on the host star photosphere \citep{Oshagh2014,Rackham2018,Mallonn2018}. Without a proper knowledge of all factors that potentially influence the measured slopes in exoplanet transmission spectra, a correct scientific interpretation is not possible. In this article, we wish to direct the attention of observers to a degeneracy of the orbital parameters with the optical exoplanet spectral slope.

The transmission spectrum of HAT-P-12b finally derived in this work has the best currently achievable precision. The data points shortward of 700~nm show a standard deviation of about 0.6 atmospheric pressure scale heights ($H$) with a mean uncertainty of about 0.8~$H,$ including the less precise data points shortward of 400~nm. This is remarkable considering that the host star is comparably faint, with a V-band magnitude of 12.8, and that it is a K-type star with rather little stellar flux at short wavelengths. Our transmission spectrum is inconsistent with the clear-atmosphere, solar-composition model by \cite{fortney2010}, scaled to the parameters of this planet (Fig. \ref{final_spectr}). There is no indication for absorption by sodium in either an HST wavelength channel centered at the line core or at the wavelength of pressure-broadened wings, as predicted by the \cite{fortney2010} model. The transmission data longward of 700~nm are less precise with larger uncertainties and larger scatter among the data points. These data did not show a significant indication for a pressure-broadened potassium line but they could not rule it out either. We agree with the finding by S16 of a tentative absorption signal in the core of the potassium line, but it is just significant at the $2\,\sigma$ confidence level and based on only one single light-curve. Further observations are required to confirm this. 

A linear regression line calculated using all our data points shows a slope of $-2.96\,\pm\,1.28\,\times\,10^{-6}$~nm$^{-1}$. We consider this slope to be a signature of the planetary atmosphere because of the low activity level of the host star \citep{Mallonn2015,Mancini2018}. In units of $H$, this slope corresponds to a difference of about 1.5~$H$ from the u' band to the z' band. With this amplitude, the spectral slope is rather typical for hot Jupiters \citep{sing2016,Mallonnwakeford2017,Pinhas2017}. HAT-P-12b has an equilibrium temperature of 950~K (Section \ref{sec_physpara}), thus, it is the coolest object of the ten planets in the HST survey. However, no dependence of the slope amplitude on temperature becomes evident, as HAT-P-12b has a very similar slope to 1100~K WASP-39b, 1450~K HD209458b, and even 2500~K WASP-12b \citep{sing2016,Mallonnwakeford2017}. HAT-P-12b is similar in equilibrium temperature and surface gravity to WASP-39b. However, the former presents an optical spectrum that is dominated by haze particles scattering, while the latter shows a rather clear atmosphere that reveals the pressure-broadened wings of sodium \citep{Fischer2016,Wakeford2018}. Future work may reveal whether this diversity is linked to differences in cloud condensation by different planetary metallicity abundances or planetary atmospheric dynamics. Another possibility may be differences in the production of photochemical haze by the different host star spectral types of HAT-P-12 and WASP-39.

\begin{acknowledgements}
We thank the anonymous referee for the insightful comments. We thank Nikolay Nikolov and David Sing for discussions and comments on the manuscript. We are grateful to Jonathan Fortney for providing atmospheric models for HAT-P-12b, and we thank Antonio Claret for providing limb-darkening coefficients of HAT-P-12. We are grateful for DDT observing time at Calar Alto and wish to acknowledge all technical support at the involved observing facilities. L.M. acknowledges support from the Italian Minister of Instruction, University and Research (MIUR) through FFABR 2017 fund. L.M. acknowledges support from the University of Rome Tor Vergata through ``Mission: Sustainability 2016'' fund. This research has made use of the SIMBAD data base and VizieR catalog access tool, operated at CDS, Strasbourg, France, and of the NASA Astrophysics Data System (ADS).
This work is partly based on data obtained with the STELLA robotic telescopes in Tenerife, an AIP facility jointly operated by AIP and IAC, partly based on observations made with the Italian Telescopio Nazionale Galileo (TNG) operated on the island of La Palma by the Fundación Galileo Galilei of the INAF (Istituto Nazionale di Astrofisica) at the Spanish Observatorio del Roque de los Muchachos of the Instituto de Astrofisica de Canarias, partly based on observations collected at the German-Spanish Astronomical Center, Calar Alto, jointly operated by the Max-Planck-Institut für Astronomie Heidelberg and the Instituto de Astrofísica de Andalucía (CSIC), and partly based on observations obtained with the Apache Point Observatory 3.5-meter telescope, which is owned and operated by the Astrophysical Research Consortium.

\end{acknowledgements}

\bibliographystyle{aa}
\bibliography{mybib}

\begin{thebibliography}{82}
\expandafter\ifx\csname natexlab\endcsname\relax\def\natexlab#1{#1}\fi

\bibitem[{{Barrado} {et~al.}(2011){Barrado}, {Thiele}, {Aceituno}, {Pedraz},
  {S{\'a}nchez}, {Aguirre}, {Alises}, {Bergond}, {Galad{\'{\i}}}, {Guijarro},
  {Hoyo}, {Mast}, {Montoya}, {Sengupta}, {de Guindos}, \& {Solano}}]{dlr-mk3}
{Barrado}, D., {Thiele}, U., {Aceituno}, J., {et~al.} 2011, in Highlights of
  Spanish Astrophysics VI, ed. M.~R. {Zapatero Osorio}, J.~{Gorgas},
  J.~{Ma{\'{\i}}z Apell{\'a}niz}, J.~R. {Pardo}, \& A.~{Gil de Paz}, 637--646

\bibitem[{{Berta} {et~al.}(2012){Berta}, {Charbonneau}, {D{\'e}sert},
  {Miller-Ricci Kempton}, {McCullough}, {Burke}, {Fortney}, {Irwin}, {Nutzman},
  \& {Homeier}}]{Berta2012}
{Berta}, Z.~K., {Charbonneau}, D., {D{\'e}sert}, J.-M., {et~al.} 2012, \apj,
  747, 35

\bibitem[{{Bertin} \& {Arnouts}(1996)}]{Bertin96}
{Bertin}, E. \& {Arnouts}, S. 1996, \aaps, 117, 393

\bibitem[{{Biddle} {et~al.}(2014){Biddle}, {Pearson}, {Crossfield}, {Fulton},
  {Ciceri}, {Eastman}, {Barman}, {Mann}, {Henry}, {Howard}, {Williamson},
  {Sinukoff}, {Dragomir}, {Vican}, {Mancini}, {Southworth}, {Greenberg},
  {Turner}, {Thompson}, {Taylor}, {Levine}, \& {Webber}}]{Biddle2014}
{Biddle}, L.~I., {Pearson}, K.~A., {Crossfield}, I.~J.~M., {et~al.} 2014,
  \mnras, 443, 1810

\bibitem[{{Burrows}(2014)}]{Burrows2014}
{Burrows}, A.~S. 2014, \nat, 513, 345

\bibitem[{{Charbonneau} {et~al.}(2002){Charbonneau}, {Brown}, {Noyes}, \&
  {Gilliland}}]{Charbonneau2002}
{Charbonneau}, D., {Brown}, T.~M., {Noyes}, R.~W., \& {Gilliland}, R.~L. 2002,
  \apj, 568, 377

\bibitem[{{Chen} {et~al.}(2017){Chen}, {Guenther}, {Pall{\'e}}, {Nortmann},
  {Nowak}, {Kunz}, {Parviainen}, \& {Murgas}}]{Chen2017}
{Chen}, G., {Guenther}, E.~W., {Pall{\'e}}, E., {et~al.} 2017, \aap, 600, A138

\bibitem[{{Claret}(2004)}]{Claret2004}
{Claret}, A. 2004, \aap, 428, 1001

\bibitem[{{Claret} \& {Bloemen}(2011)}]{Claret2011}
{Claret}, A. \& {Bloemen}, S. 2011, \aap, 529, A75

\bibitem[{{Csizmadia} {et~al.}(2013){Csizmadia}, {Pasternacki}, {Dreyer},
  {Cabrera}, {Erikson}, \& {Rauer}}]{Csizmadia2013}
{Csizmadia}, S., {Pasternacki}, T., {Dreyer}, C., {et~al.} 2013, \aap, 549, A9

\bibitem[{{Deming} \& {Seager}(2017)}]{Deming2017}
{Deming}, D. \& {Seager}, S. 2017, ArXiv e-prints

\bibitem[{{Dragomir} {et~al.}(2015){Dragomir}, {Benneke}, {Pearson},
  {Crossfield}, {Eastman}, {Barman}, \& {Biddle}}]{Dragomir2015}
{Dragomir}, D., {Benneke}, B., {Pearson}, K.~A., {et~al.} 2015, \apj, 814, 102

\bibitem[{{Fazio} {et~al.}(2004){Fazio}, {Hora}, {Allen}, {Ashby}, {Barmby},
  {Deutsch}, {Huang}, {Kleiner}, {Marengo}, {Megeath}, {Melnick}, {Pahre},
  {Patten}, {Polizotti}, {Smith}, {Taylor}, {Wang}, {Willner}, {Hoffmann},
  {Pipher}, {Forrest}, {McMurty}, {McCreight}, {McKelvey}, {McMurray}, {Koch},
  {Moseley}, {Arendt}, {Mentzell}, {Marx}, {Losch}, {Mayman}, {Eichhorn},
  {Krebs}, {Jhabvala}, {Gezari}, {Fixsen}, {Flores}, {Shakoorzadeh}, {Jungo},
  {Hakun}, {Workman}, {Karpati}, {Kichak}, {Whitley}, {Mann}, {Tollestrup},
  {Eisenhardt}, {Stern}, {Gorjian}, {Bhattacharya}, {Carey}, {Nelson},
  {Glaccum}, {Lacy}, {Lowrance}, {Laine}, {Reach}, {Stauffer}, {Surace},
  {Wilson}, {Wright}, {Hoffman}, {Domingo}, \& {Cohen}}]{irac}
{Fazio}, G.~G., {Hora}, J.~L., {Allen}, L.~E., {et~al.} 2004, \apjs, 154, 10

\bibitem[{{Fischer} {et~al.}(2016){Fischer}, {Knutson}, {Sing}, {Henry},
  {Williamson}, {Fortney}, {Burrows}, {Kataria}, {Nikolov}, {Showman},
  {Ballester}, {D{\'e}sert}, {Aigrain}, {Deming}, {Lecavelier des Etangs}, \&
  {Vidal-Madjar}}]{Fischer2016}
{Fischer}, P.~D., {Knutson}, H.~A., {Sing}, D.~K., {et~al.} 2016, \apj, 827, 19

\bibitem[{{Fortney} {et~al.}(2010){Fortney}, {Shabram}, {Showman}, {Lian},
  {Freedman}, {Marley}, \& {Lewis}}]{fortney2010}
{Fortney}, J.~J., {Shabram}, M., {Showman}, A.~P., {et~al.} 2010, \apj, 709,
  1396

\bibitem[{{Gibson} {et~al.}(2017){Gibson}, {Nikolov}, {Sing}, {Barstow},
  {Evans}, {Kataria}, \& {Wilson}}]{Gibson2017}
{Gibson}, N.~P., {Nikolov}, N., {Sing}, D.~K., {et~al.} 2017, \mnras, 467, 4591

\bibitem[{{Hartman} {et~al.}(2009){Hartman}, {Bakos}, {Torres}, {Kov{\'a}cs},
  {Noyes}, {P{\'a}l}, {Latham}, {Sip{\H o}cz}, {Fischer}, {Johnson}, {Marcy},
  {Butler}, {Howard}, {Esquerdo}, {Sasselov}, {Kov{\'a}cs}, {Stefanik},
  {Fernandez}, {L{\'a}z{\'a}r}, {Papp}, \& {S{\'a}ri}}]{Hartman2009}
{Hartman}, J.~D., {Bakos}, G.~{\'A}., {Torres}, G., {et~al.} 2009, \apj, 706,
  785

\bibitem[{{Huehnerhoff} {et~al.}(2016){Huehnerhoff}, {Ketzeback}, {Bradley},
  {Dembicky}, {Doughty}, {Hawley}, {Johnson}, {Klaene}, {Leon}, {McMillan},
  {Owen}, {Sayres}, {Sheen}, \& {Shugart}}]{arctic}
{Huehnerhoff}, J., {Ketzeback}, W., {Bradley}, A., {et~al.} 2016, in \procspie,
  Vol. 9908, Ground-based and Airborne Instrumentation for Astronomy VI, 99085H

\bibitem[{{Huitson} {et~al.}(2013){Huitson}, {Sing}, {Pont}, {Fortney},
  {Burrows}, {Wilson}, {Ballester}, {Nikolov}, {Gibson}, {Deming}, {Aigrain},
  {Evans}, {Henry}, {Lecavelier des Etangs}, {Showman}, {Vidal-Madjar}, \&
  {Zahnle}}]{Huitson2013}
{Huitson}, C.~M., {Sing}, D.~K., {Pont}, F., {et~al.} 2013, \mnras, 434, 3252

\bibitem[{{Jord{\'a}n} {et~al.}(2013){Jord{\'a}n}, {Espinoza}, {Rabus},
  {Eyheramendy}, {Sing}, {D{\'e}sert}, {Bakos}, {Fortney}, {L{\'o}pez-Morales},
  {Maxted}, {Triaud}, \& {Szentgyorgyi}}]{Jordan2013}
{Jord{\'a}n}, A., {Espinoza}, N., {Rabus}, M., {et~al.} 2013, \apj, 778, 184

\bibitem[{{Juvan} {et~al.}(2018){Juvan}, {Lendl}, {Cubillos}, {Fossati},
  {Tregloan-Reed}, {Lammer}, {Guenther}, \& {Hanslmeier}}]{Juvan2018}
{Juvan}, I.~G., {Lendl}, M., {Cubillos}, P.~E., {et~al.} 2018, \aap, 610, A15

\bibitem[{{Kirk} {et~al.}(2018){Kirk}, {Wheatley}, {Louden}, {Skillen}, {King},
  {McCormac}, \& {Irwin}}]{Kirk2018}
{Kirk}, J., {Wheatley}, P.~J., {Louden}, T., {et~al.} 2018, \mnras, 474, 876

\bibitem[{{Knutson} {et~al.}(2014){Knutson}, {Fulton}, {Montet}, {Kao}, {Ngo},
  {Howard}, {Crepp}, {Hinkley}, {Bakos}, {Batygin}, {Johnson}, {Morton}, \&
  {Muirhead}}]{Knutson2014}
{Knutson}, H.~A., {Fulton}, B.~J., {Montet}, B.~T., {et~al.} 2014, \apj, 785,
  126

\bibitem[{{Kreidberg} {et~al.}(2014){Kreidberg}, {Bean}, {D{\'e}sert},
  {Benneke}, {Deming}, {Stevenson}, {Seager}, {Berta-Thompson}, {Seifahrt}, \&
  {Homeier}}]{Kreidberg2014}
{Kreidberg}, L., {Bean}, J.~L., {D{\'e}sert}, J.-M., {et~al.} 2014, \nat, 505,
  69

\bibitem[{{Kreidberg} {et~al.}(2018){Kreidberg}, {Line}, {Thorngren}, {Morley},
  \& {Stevenson}}]{Kreidberg2018}
{Kreidberg}, L., {Line}, M.~R., {Thorngren}, D., {Morley}, C.~V., \&
  {Stevenson}, K.~B. 2018, \apjl, 858, L6

\bibitem[{{Leckrone} {et~al.}(1998){Leckrone}, {Cheng}, {Feinberg}, {Trauger},
  {Macchetto}, \& {MacKenty}}]{wfc3}
{Leckrone}, D.~S., {Cheng}, E.~S., {Feinberg}, L.~D., {et~al.} 1998, in
  Bulletin of the American Astronomical Society, Vol.~30, American Astronomical
  Society Meeting Abstracts \#192, 861

\bibitem[{{Lee} {et~al.}(2012){Lee}, {Youn}, {Kim}, {Lee}, \&
  {Hinse}}]{Lee2012}
{Lee}, J.~W., {Youn}, J.-H., {Kim}, S.-L., {Lee}, C.-U., \& {Hinse}, T.~C.
  2012, \aj, 143, 95

\bibitem[{{Lendl} {et~al.}(2016){Lendl}, {Delrez}, {Gillon}, {Madhusudhan},
  {Jehin}, {Queloz}, {Anderson}, {Demory}, \& {Hellier}}]{Lendl2016}
{Lendl}, M., {Delrez}, L., {Gillon}, M., {et~al.} 2016, \aap, 587, A67

\bibitem[{{Line} {et~al.}(2013){Line}, {Knutson}, {Deming}, {Wilkins}, \&
  {Desert}}]{Line2013}
{Line}, M.~R., {Knutson}, H., {Deming}, D., {Wilkins}, A., \& {Desert}, J.-M.
  2013, \apj, 778, 183

\bibitem[{{Maciejewski} {et~al.}(2015){Maciejewski}, {Fern{\'a}ndez},
  {Aceituno}, {Ohlert}, {Puchalski}, {Dimitrov}, {Seeliger}, {Kitze}, {Raetz},
  {Errmann}, {Gilbert}, {Pannicke}, {Schmidt}, \&
  {Neuh{\"a}user}}]{Maciejewski2015}
{Maciejewski}, G., {Fern{\'a}ndez}, M., {Aceituno}, F.~J., {et~al.} 2015, \aap,
  577, A109

\bibitem[{{Mackebrandt} {et~al.}(2017){Mackebrandt}, {Mallonn}, {Ohlert},
  {Granzer}, {Lalitha}, {Garc{\'{\i}}a Mu{\~n}oz}, {Gibson}, {Lee}, {Sozzetti},
  {Turner}, {Va{\v n}ko}, \& {Strassmeier}}]{Mackebrandt2017}
{Mackebrandt}, F., {Mallonn}, M., {Ohlert}, J.~M., {et~al.} 2017, \aap, 608,
  A26

\bibitem[{{Madhusudhan} {et~al.}(2016){Madhusudhan}, {Ag{\'u}ndez}, {Moses}, \&
  {Hu}}]{Madhusudhan2016}
{Madhusudhan}, N., {Ag{\'u}ndez}, M., {Moses}, J.~I., \& {Hu}, Y. 2016, \ssr,
  205, 285

\bibitem[{{Mallonn} {et~al.}(2016){Mallonn}, {Bernt}, {Herrero}, {Hoyer},
  {Kirk}, {Wheatley}, {Seeliger}, {Mackebrandt}, {von Essen}, {Strassmeier},
  {Granzer}, {K{\"u}nstler}, {Dhillon}, {Marsh}, \& {Gaitan}}]{Mallonn2016}
{Mallonn}, M., {Bernt}, I., {Herrero}, E., {et~al.} 2016, \mnras, 463, 604

\bibitem[{{Mallonn} {et~al.}(2018){Mallonn}, {Herrero}, {Juvan}, {von Essen},
  {Rosich}, {Ribas}, {Granzer}, {Alexoudi}, \& {Strassmeier}}]{Mallonn2018}
{Mallonn}, M., {Herrero}, E., {Juvan}, I.~G., {et~al.} 2018, ArXiv e-prints

\bibitem[{{Mallonn} {et~al.}(2015){Mallonn}, {Nascimbeni}, {Weingrill}, {von
  Essen}, {Strassmeier}, {Piotto}, {Pagano}, {Scandariato}, {Csizmadia},
  {Herrero}, {Sada}, {Dhillon}, {Marsh}, {K{\"u}nstler}, {Bernt}, \&
  {Granzer}}]{Mallonn2015}
{Mallonn}, M., {Nascimbeni}, V., {Weingrill}, J., {et~al.} 2015, \aap, 583,
  A138

\bibitem[{{Mallonn} \& {Wakeford}(2017)}]{Mallonnwakeford2017}
{Mallonn}, M. \& {Wakeford}, H.~R. 2017, Astronomische Nachrichten, 338, 773

\bibitem[{{Mancini} {et~al.}(2018){Mancini}, {Esposito}, {Covino},
  {Southworth}, {Biazzo}, {Bruni}, {Ciceri}, {Evans}, {Lanza}, {Poretti},
  {Sarkis}, {Smith}, {Brogi}, {Affer}, {Benatti}, {Bignamini}, {Boccato},
  {Bonomo}, {Borsa}, {Carleo}, {Claudi}, {Cosentino}, {Damasso}, {Desidera},
  {Giacobbe}, {Gonz{\'a}lez-{\'A}lvarez}, {Gratton}, {Harutyunyan}, {Leto},
  {Maggio}, {Malavolta}, {Maldonado}, {Martinez-Fiorenzano}, {Masiero},
  {Micela}, {Molinari}, {Nascimbeni}, {Pagano}, {Pedani}, {Piotto}, {Rainer},
  {Scandariato}, {Smareglia}, {Sozzetti}, {Andreuzzi}, \&
  {Henning}}]{Mancini2018}
{Mancini}, L., {Esposito}, M., {Covino}, E., {et~al.} 2018, \aap, 613, A41

\bibitem[{{Mancini} {et~al.}(2016){Mancini}, {Giordano}, {Molli{\`e}re},
  {Southworth}, {Brahm}, {Ciceri}, \& {Henning}}]{Mancini2016}
{Mancini}, L., {Giordano}, M., {Molli{\`e}re}, P., {et~al.} 2016, \mnras, 461,
  1053

\bibitem[{{Mandel} \& {Agol}(2002)}]{Mandel&Agol2002}
{Mandel}, K. \& {Agol}, E. 2002, \apjl, 580, L171

\bibitem[{{Mayor} \& {Queloz}(1995)}]{Mayor&Queloz1995}
{Mayor}, M. \& {Queloz}, D. 1995, \nat, 378, 355

\bibitem[{{Meisenheimer}(1994)}]{cafos}
{Meisenheimer}, K. 1994, Sterne und Weltraum, 33, 516

\bibitem[{{Nascimbeni} {et~al.}(2015){Nascimbeni}, {Mallonn}, {Scandariato},
  {Pagano}, {Piotto}, {Micela}, {Messina}, {Leto}, {Strassmeier}, {Bisogni}, \&
  {Speziali}}]{Nascimbeni2015}
{Nascimbeni}, V., {Mallonn}, M., {Scandariato}, G., {et~al.} 2015, \aap, 579,
  A113

\bibitem[{{Nascimbeni} {et~al.}(2013){Nascimbeni}, {Piotto}, {Pagano},
  {Scandariato}, {Sani}, \& {Fumana}}]{Nascimbeni2013}
{Nascimbeni}, V., {Piotto}, G., {Pagano}, I., {et~al.} 2013, \aap, 559, A32

\bibitem[{{Nikolov} {et~al.}(2015){Nikolov}, {Sing}, {Burrows}, {Fortney},
  {Henry}, {Pont}, {Ballester}, {Aigrain}, {Wilson}, {Huitson}, {Gibson},
  {D{\'e}sert}, {Lecavelier Des Etangs}, {Showman}, {Vidal-Madjar}, {Wakeford},
  \& {Zahnle}}]{Nikolov2015}
{Nikolov}, N., {Sing}, D.~K., {Burrows}, A.~S., {et~al.} 2015, \mnras, 447, 463

\bibitem[{{Nikolov} {et~al.}(2018){Nikolov}, {Sing}, {Fortney}, {Goyal},
  {Drummond}, {Evans}, {Gibson}, {De Mooij}, {Rustamkulov}, {Wakeford},
  {Smalley}, {Burgasser}, {Helling}, {Mayne}, {Madhusudhan}, {Kataria},
  {Baines}, {Carter}, {Ballester}, {Barstow}, {McCleery}, \&
  {Spake}}]{Nikolov2018}
{Nikolov}, N., {Sing}, D.~K., {Fortney}, J.~J., {et~al.} 2018, \nat, 557, 526

\bibitem[{{Nikolov} {et~al.}(2016){Nikolov}, {Sing}, {Gibson}, {Fortney},
  {Evans}, {Barstow}, {Kataria}, \& {Wilson}}]{nikolov2016}
{Nikolov}, N., {Sing}, D.~K., {Gibson}, N.~P., {et~al.} 2016, \apj, 832, 191

\bibitem[{{Nikolov} {et~al.}(2014){Nikolov}, {Sing}, {Pont}, {Burrows},
  {Fortney}, {Ballester}, {Evans}, {Huitson}, {Wakeford}, {Wilson}, {Aigrain},
  {Deming}, {Gibson}, {Henry}, {Knutson}, {Lecavelier des Etangs}, {Showman},
  {Vidal-Madjar}, \& {Zahnle}}]{Nikolov2014}
{Nikolov}, N., {Sing}, D.~K., {Pont}, F., {et~al.} 2014, \mnras, 437, 46

\bibitem[{{Oliva}(2006)}]{dolores}
{Oliva}, E. 2006, Memorie della Societa Astronomica Italiana Supplementi, 9,
  409

\bibitem[{{Oshagh} {et~al.}(2014){Oshagh}, {Santos}, {Ehrenreich},
  {Haghighipour}, {Figueira}, {Santerne}, \& {Montalto}}]{Oshagh2014}
{Oshagh}, M., {Santos}, N.~C., {Ehrenreich}, D., {et~al.} 2014, \aap, 568, A99

\bibitem[{{Parviainen} {et~al.}(2018){Parviainen}, {Pall{\'e}}, {Chen},
  {Nortmann}, {Murgas}, {Nowak}, {Aigrain}, {Booth}, {Abazorius}, \&
  {Iro}}]{Parviainen2018}
{Parviainen}, H., {Pall{\'e}}, E., {Chen}, G., {et~al.} 2018, \aap, 609, A33

\bibitem[{{Parviainen} {et~al.}(2016){Parviainen}, {Pall{\'e}}, {Nortmann},
  {Nowak}, {Iro}, {Murgas}, \& {Aigrain}}]{Parviainen2016}
{Parviainen}, H., {Pall{\'e}}, E., {Nortmann}, L., {et~al.} 2016, \aap, 585,
  A114

\bibitem[{{Pearson} {et~al.}(2014){Pearson}, {Turner}, \&
  {Sagan}}]{Pearson2014}
{Pearson}, K.~A., {Turner}, J.~D., \& {Sagan}, T.~G. 2014, \na, 27, 102

\bibitem[{{Pinhas} \& {Madhusudhan}(2017)}]{Pinhas2017}
{Pinhas}, A. \& {Madhusudhan}, N. 2017, \mnras, 471, 4355

\bibitem[{{Pont} {et~al.}(2013){Pont}, {Sing}, {Gibson}, {Aigrain}, {Henry}, \&
  {Husnoo}}]{Pont2013}
{Pont}, F., {Sing}, D.~K., {Gibson}, N.~P., {et~al.} 2013, \mnras, 432, 2917

\bibitem[{{Pont} {et~al.}(2006){Pont}, {Zucker}, \& {Queloz}}]{Pont2006}
{Pont}, F., {Zucker}, S., \& {Queloz}, D. 2006, \mnras, 373, 231

\bibitem[{{Rackham} {et~al.}(2018){Rackham}, {Apai}, \&
  {Giampapa}}]{Rackham2018}
{Rackham}, B.~V., {Apai}, D., \& {Giampapa}, M.~S. 2018, \apj, 853, 122

\bibitem[{{Sada} {et~al.}(2012){Sada}, {Deming}, {Jennings}, {Jackson},
  {Hamilton}, {Fraine}, {Peterson}, {Haase}, {Bays}, {Lunsford}, \&
  {O'Gorman}}]{Sada2012}
{Sada}, P.~V., {Deming}, D., {Jennings}, D.~E., {et~al.} 2012, \pasp, 124, 212

\bibitem[{{Safronov}(1972)}]{Safronov1972}
{Safronov}, V.~S. 1972, {Evolution of the protoplanetary cloud and formation of
  the earth and planets.}

\bibitem[{Schwarz(1978)}]{Schwarz1978}
Schwarz, G. 1978, The Annals of Statistics, 6, 461

\bibitem[{{Seager}(2011)}]{Seager2011}
{Seager}, S. 2011, in IAU Symposium, Vol. 276, The Astrophysics of Planetary
  Systems: Formation, Structure, and Dynamical Evolution, ed. A.~{Sozzetti},
  M.~G. {Lattanzi}, \& A.~P. {Boss}, 198--207

\bibitem[{{Sedaghati} {et~al.}(2017){Sedaghati}, {Boffin}, {Delrez}, {Gillon},
  {Csizmadia}, {Smith}, \& {Rauer}}]{Sedaghati2017}
{Sedaghati}, E., {Boffin}, H.~M.~J., {Delrez}, L., {et~al.} 2017, \mnras, 468,
  3123

\bibitem[{{Sing}(2018)}]{Sing2018}
{Sing}, D.~K. 2018, ArXiv e-prints

\bibitem[{{Sing} {et~al.}(2016){Sing}, {Fortney}, {Nikolov}, {Wakeford},
  {Kataria}, {Evans}, {Aigrain}, {Ballester}, {Burrows}, {Deming},
  {D{\'e}sert}, {Gibson}, {Henry}, {Huitson}, {Knutson}, {Lecavelier Des
  Etangs}, {Pont}, {Showman}, {Vidal-Madjar}, {Williamson}, \&
  {Wilson}}]{sing2016}
{Sing}, D.~K., {Fortney}, J.~J., {Nikolov}, N., {et~al.} 2016, \nat, 529, 59

\bibitem[{{Sing} {et~al.}(2013){Sing}, {Lecavelier des Etangs}, {Fortney},
  {Burrows}, {Pont}, {Wakeford}, {Ballester}, {Nikolov}, {Henry}, {Aigrain},
  {Deming}, {Evans}, {Gibson}, {Huitson}, {Knutson}, {Showman}, {Vidal-Madjar},
  {Wilson}, {Williamson}, \& {Zahnle}}]{Sing2013}
{Sing}, D.~K., {Lecavelier des Etangs}, A., {Fortney}, J.~J., {et~al.} 2013,
  \mnras, 436, 2956

\bibitem[{{Sing} {et~al.}(2011){Sing}, {Pont}, {Aigrain}, {Charbonneau},
  {D{\'e}sert}, {Gibson}, {Gilliland}, {Hayek}, {Henry}, {Knutson}, {Lecavelier
  Des Etangs}, {Mazeh}, \& {Shporer}}]{Sing2011b}
{Sing}, D.~K., {Pont}, F., {Aigrain}, S., {et~al.} 2011, \mnras, 416, 1443

\bibitem[{{Sing} {et~al.}(2015){Sing}, {Wakeford}, {Showman}, {Nikolov},
  {Fortney}, {Burrows}, {Ballester}, {Deming}, {Aigrain}, {D{\'e}sert},
  {Gibson}, {Henry}, {Knutson}, {Lecavelier des Etangs}, {Pont},
  {Vidal-Madjar}, {Williamson}, \& {Wilson}}]{Sing2015}
{Sing}, D.~K., {Wakeford}, H.~R., {Showman}, A.~P., {et~al.} 2015, \mnras, 446,
  2428

\bibitem[{{Southworth}(2008)}]{Southworth08}
{Southworth}, J. 2008, \mnras, 386, 1644

\bibitem[{{Southworth}(2010)}]{South2010}
{Southworth}, J. 2010, \mnras, 408, 1689

\bibitem[{{Southworth} {et~al.}(2010){Southworth}, {Mancini}, {Novati},
  {Dominik}, {Glitrup}, {Hinse}, {J{\o}rgensen}, {Mathiasen}, {Ricci}, {Maier},
  {Zimmer}, {Bozza}, {Browne}, {Bruni}, {Burgdorf}, {Dall'Ora}, {Finet},
  {Harps{\o}e}, {Hundertmark}, {Liebig}, {Rahvar}, {Scarpetta}, {Skottfelt},
  {Smalley}, {Snodgrass}, \& {Surdej}}]{Southworth2010}
{Southworth}, J., {Mancini}, L., {Novati}, S.~C., {et~al.} 2010, \mnras, 408,
  1680

\bibitem[{{Southworth} {et~al.}(2007){Southworth}, {Wheatley}, \&
  {Sams}}]{Southworth2007}
{Southworth}, J., {Wheatley}, P.~J., \& {Sams}, G. 2007, \mnras, 379, L11

\bibitem[{{Strassmeier} {et~al.}(2004){Strassmeier}, {Granzer}, {Weber},
  {Woche}, {Andersen}, {Bartus}, {Bauer}, {Dionies}, {Popow}, {Fechner},
  {Hildebrandt}, {Washuettl}, {Ritter}, {Schwope}, {Staude}, {Paschke},
  {Stolz}, {Serre-Ricart}, {de la Rosa}, \& {Arnay}}]{Strassmeier2004}
{Strassmeier}, K.~G., {Granzer}, T., {Weber}, M., {et~al.} 2004, Astronomische
  Nachrichten, 325, 527

\bibitem[{{Todorov} {et~al.}(2013){Todorov}, {Deming}, {Knutson}, {Burrows},
  {Fortney}, {Lewis}, {Cowan}, {Agol}, {Desert}, {Sada}, {Charbonneau},
  {Laughlin}, {Langton}, \& {Showman}}]{Todorov2013}
{Todorov}, K.~O., {Deming}, D., {Knutson}, H.~A., {et~al.} 2013, \apj, 770, 102

\bibitem[{{Tsiaras} {et~al.}(2018){Tsiaras}, {Waldmann}, {Zingales},
  {Rocchetto}, {Morello}, {Damiano}, {Karpouzas}, {Tinetti}, {McKemmish},
  {Tennyson}, \& {Yurchenko}}]{Tsiaras2018}
{Tsiaras}, A., {Waldmann}, I.~P., {Zingales}, T., {et~al.} 2018, \aj, 155, 156

\bibitem[{{Turner} {et~al.}(2017){Turner}, {Leiter}, {Biddle}, {Pearson},
  {Hardegree-Ullman}, {Thompson}, {Teske}, {Cates}, {Cook}, {Berube},
  {Nieberding}, {Jones}, {Raphael}, {Wallace}, {Watson}, \&
  {Johnson}}]{Turner2017}
{Turner}, J.~D., {Leiter}, R.~M., {Biddle}, L.~I., {et~al.} 2017, \mnras, 472,
  3871

\bibitem[{{Turner} {et~al.}(2016){Turner}, {Pearson}, {Biddle}, {Smart},
  {Zellem}, {Teske}, {Hardegree-Ullman}, {Griffith}, {Leiter}, {Cates},
  {Nieberding}, {Smith}, {Thompson}, {Hofmann}, {Berube}, {Nguyen}, {Small},
  {Guvenen}, {Richardson}, {McGraw}, {Raphael}, {Crawford}, {Robertson},
  {Tombleson}, {Carleton}, {Towner}, {Walker-LaFollette}, {Hume}, {Watson},
  {Jones}, {Lichtenberger}, {Hoglund}, {Cook}, {Crossen}, {Jorgensen},
  {Romine}, {Thompson}, {Villegas}, {Wilson}, {Sanford}, {Taylor}, \&
  {Henz}}]{Turner2016}
{Turner}, J.~D., {Pearson}, K.~A., {Biddle}, L.~I., {et~al.} 2016, \mnras, 459,
  789

\bibitem[{{von Essen} {et~al.}(2016){von Essen}, {Cellone}, {Mallonn},
  {Tingley}, \& {Marcussen}}]{vonEssen2016}
{von Essen}, C., {Cellone}, S., {Mallonn}, M., {Tingley}, B., \& {Marcussen},
  M. 2016, ArXiv e-prints

\bibitem[{{von Essen} {et~al.}(2013){von Essen}, {Schr{\"o}ter}, {Agol}, \&
  {Schmitt}}]{vonEssen2013}
{von Essen}, C., {Schr{\"o}ter}, S., {Agol}, E., \& {Schmitt}, J.~H.~M.~M.
  2013, \aap, 555, A92

\bibitem[{{Wakeford} {et~al.}(2018){Wakeford}, {Sing}, {Deming}, {Lewis},
  {Goyal}, {Wilson}, {Barstow}, {Kataria}, {Drummond}, {Evans}, {Carter},
  {Nikolov}, {Knutson}, {Ballester}, \& {Mandell}}]{Wakeford2018}
{Wakeford}, H.~R., {Sing}, D.~K., {Deming}, D., {et~al.} 2018, \aj, 155, 29

\bibitem[{{Wakeford} {et~al.}(2017){Wakeford}, {Sing}, {Kataria}, {Deming},
  {Nikolov}, {Lopez}, {Tremblin}, {Amundsen}, {Lewis}, {Mandell}, {Fortney},
  {Knutson}, {Benneke}, \& {Evans}}]{wakeford2017}
{Wakeford}, H.~R., {Sing}, D.~K., {Kataria}, T., {et~al.} 2017, Science, 356,
  628

\bibitem[{{Winn}(2010)}]{Winn10}
{Winn}, J.~N. 2010, {Exoplanet Transits and Occultations}, ed. S.~{Seager}
  (University of Arizona Press), 55--77

\bibitem[{{Winn} {et~al.}(2008){Winn}, {Holman}, {Torres}, {McCullough},
  {Johns-Krull}, {Latham}, {Shporer}, {Mazeh}, {Garcia-Melendo}, {Foote},
  {Esquerdo}, \& {Everett}}]{Winn08}
{Winn}, J.~N., {Holman}, M.~J., {Torres}, G., {et~al.} 2008, \apj, 683, 1076

\bibitem[{{Woodgate} {et~al.}(1997){Woodgate}, {Kimble}, {Bowers}, {Kraemer},
  {Kaiser}, {Gull}, {Danks}, {Grady}, {Loiacono}, {Brumfield}, {Feinberg},
  {Hood}, {Meyer}, {Vanhouten}, {Argabright}, {Bybee}, {Timothy}, {Blouke},
  {Dorn}, {Bottema}, {Woodruff}, {Michika}, {Sullivan}, {Hetlinger}, {Stocker},
  {Ludtke}, {Ebbets}, {Delamere}, {Rose}, {Gardner}, {Breyer}, {Lindler},
  {Content}, {Standley}, {Hartig}, {Heap}, {Joseph}, {Green}, {Jenkins},
  {Linsky}, {Hutchings}, {Moos}, {Boggess}, {Maran}, {Roesler}, \&
  {Weistrop}}]{hst-stis}
{Woodgate}, B., {Kimble}, R., {Bowers}, C., {et~al.} 1997, in Bulletin of the
  American Astronomical Society, Vol.~29, American Astronomical Society Meeting
  Abstracts \#190, 836

\end{thebibliography}

\begin{appendix}
\section{Additional figures}

\begin{figure*}[h]
\centering
\includegraphics[width=20cm]{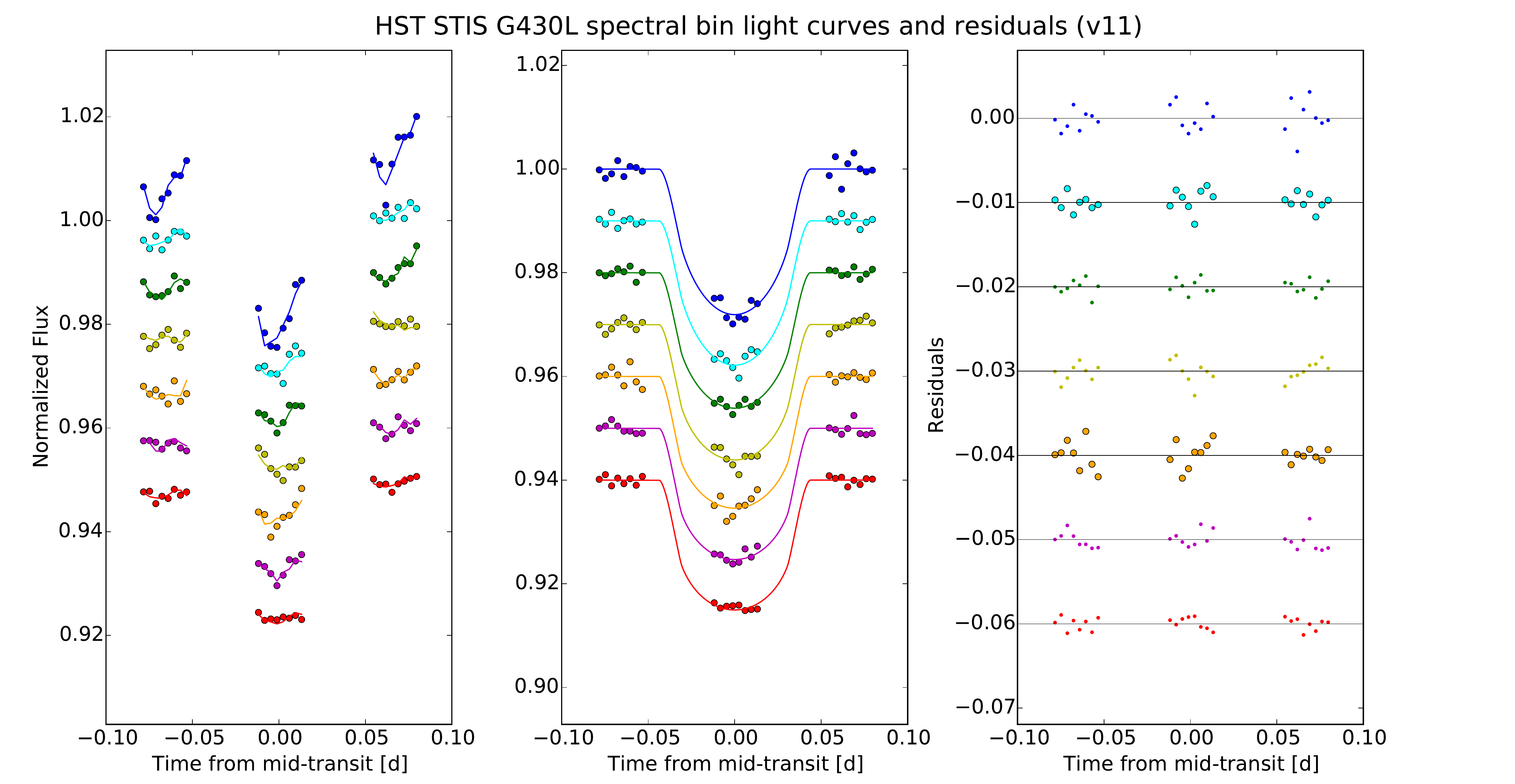}
\caption{Raw light curves of HST visit 11 in the left panel and the detrended light curves in the middle panel. Overplotted are the fitted models of the homogeneous analysis of Section \ref{sec_homoreana}. The associated residuals are presented in the right panel. The curves are shifted arbitrarily for clarity.}
\label{hst_i88a}
\end{figure*}

\begin{figure*}[h]
\centering
\includegraphics[width=20cm]{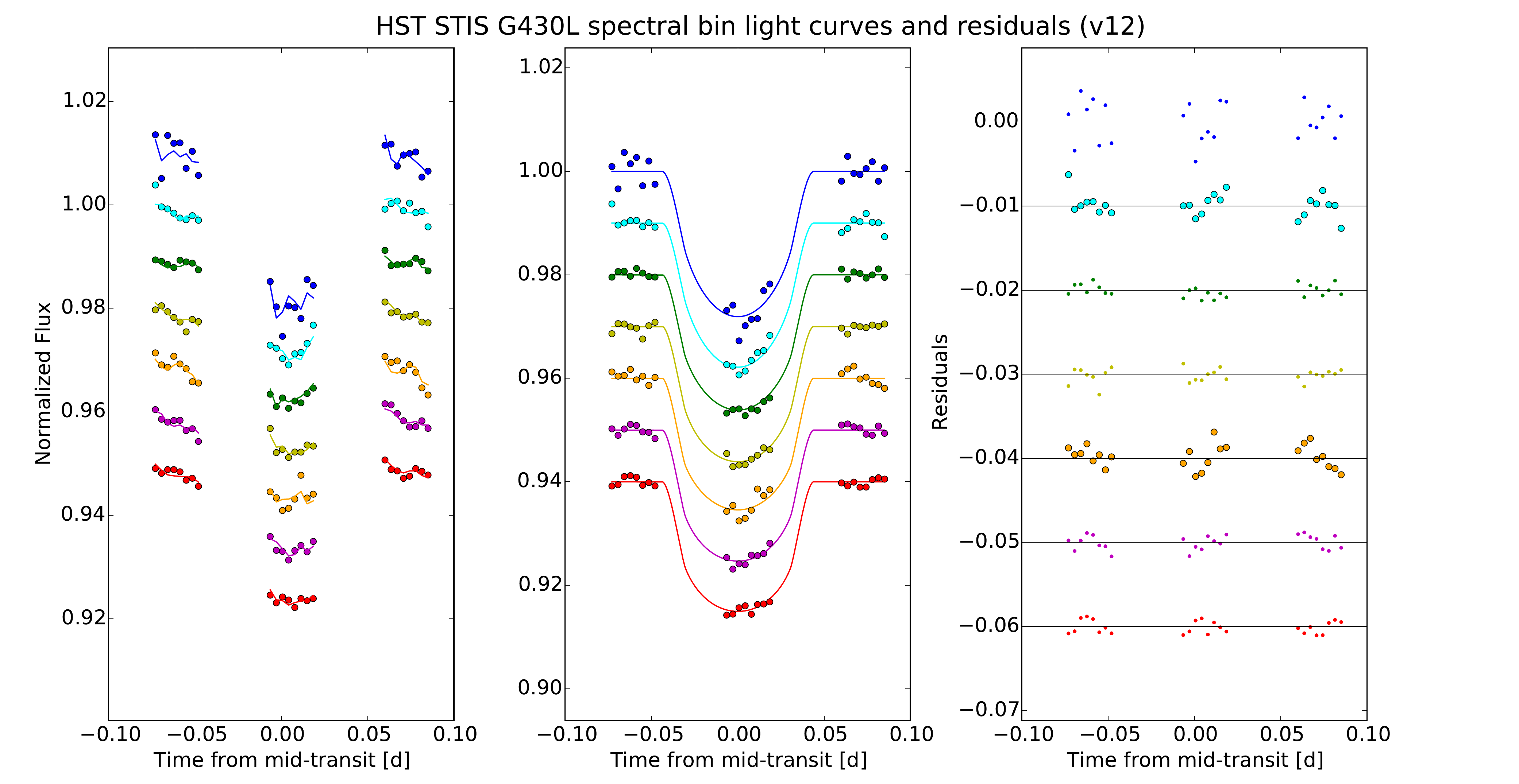}
\includegraphics[width=20cm]{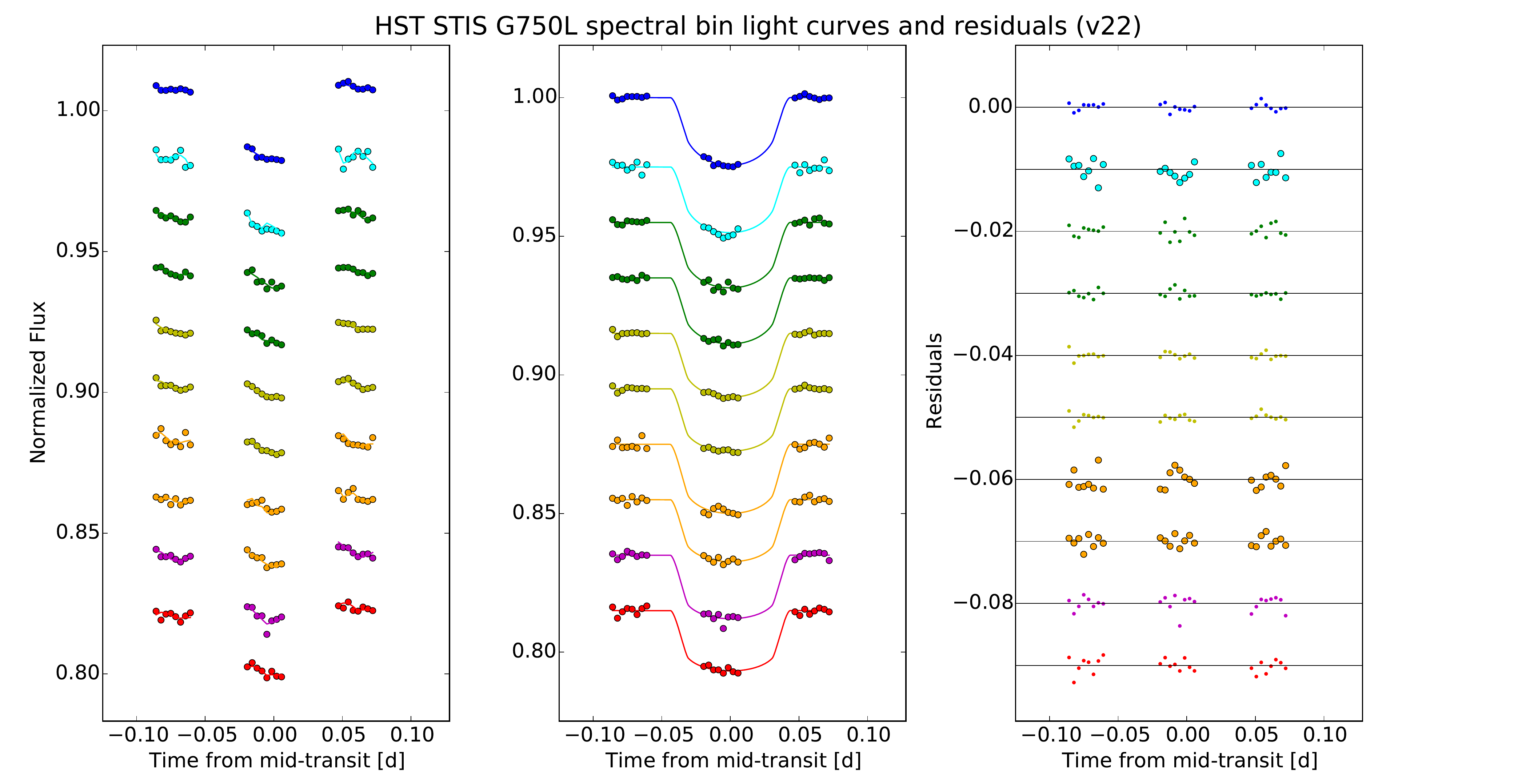}
\caption{Same as Figure \ref{hst_i88a} for HST visit 12 (upper panel) and HST visit 22 (lower panel). }
\label{hst_i88c}
\end{figure*}

\begin{figure*}[h]
\centering
\includegraphics[height=11cm]{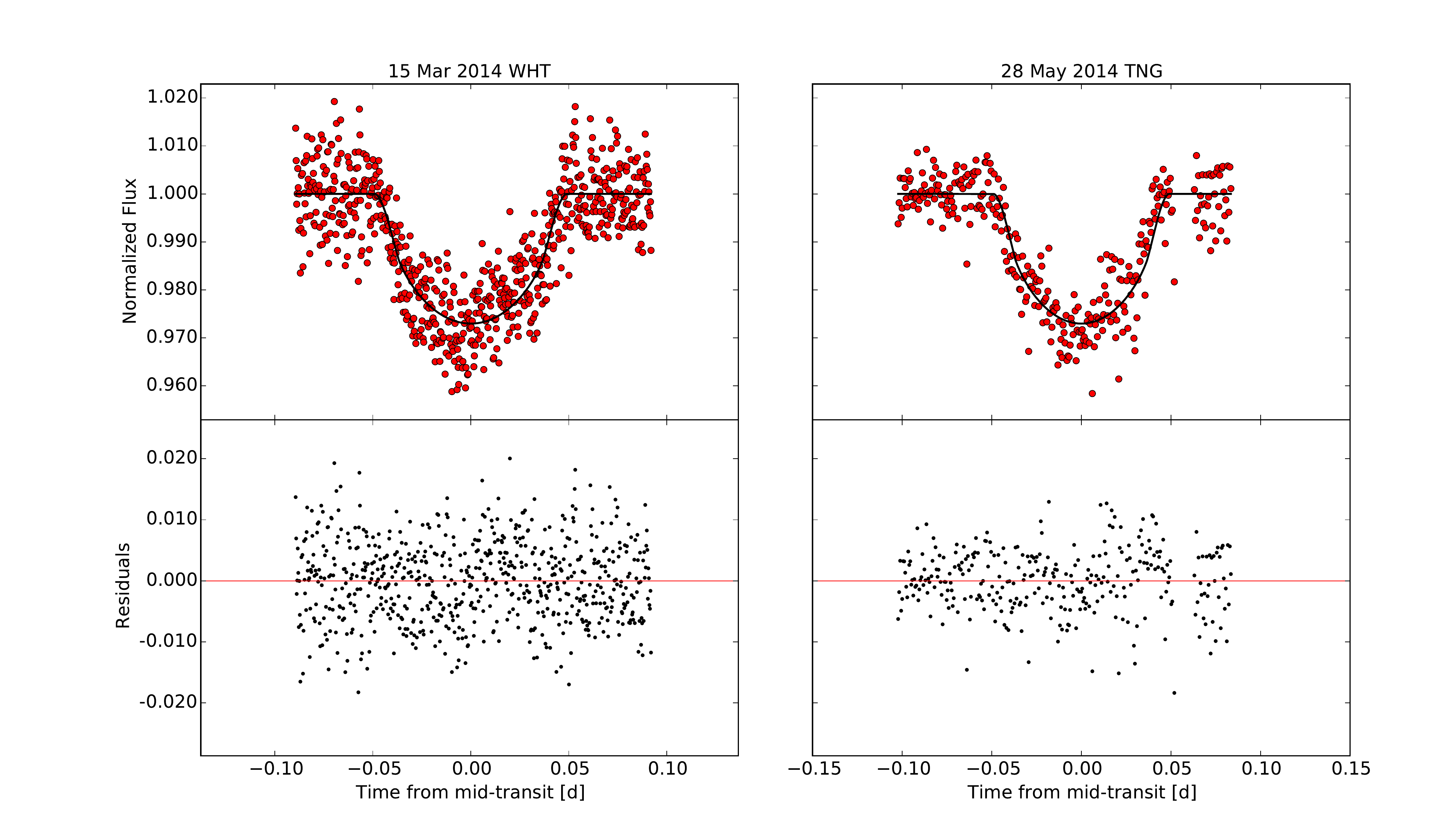}
\caption{Reanalyzed, detrended literature u'-band light curves with the associated transit model described in Section \ref{sec_homoreana}. The bottom panel shows the associated residuals.}
\label{u_band_final}
\end{figure*}

\begin{figure*}[h]
\centering
\includegraphics[height=11cm]{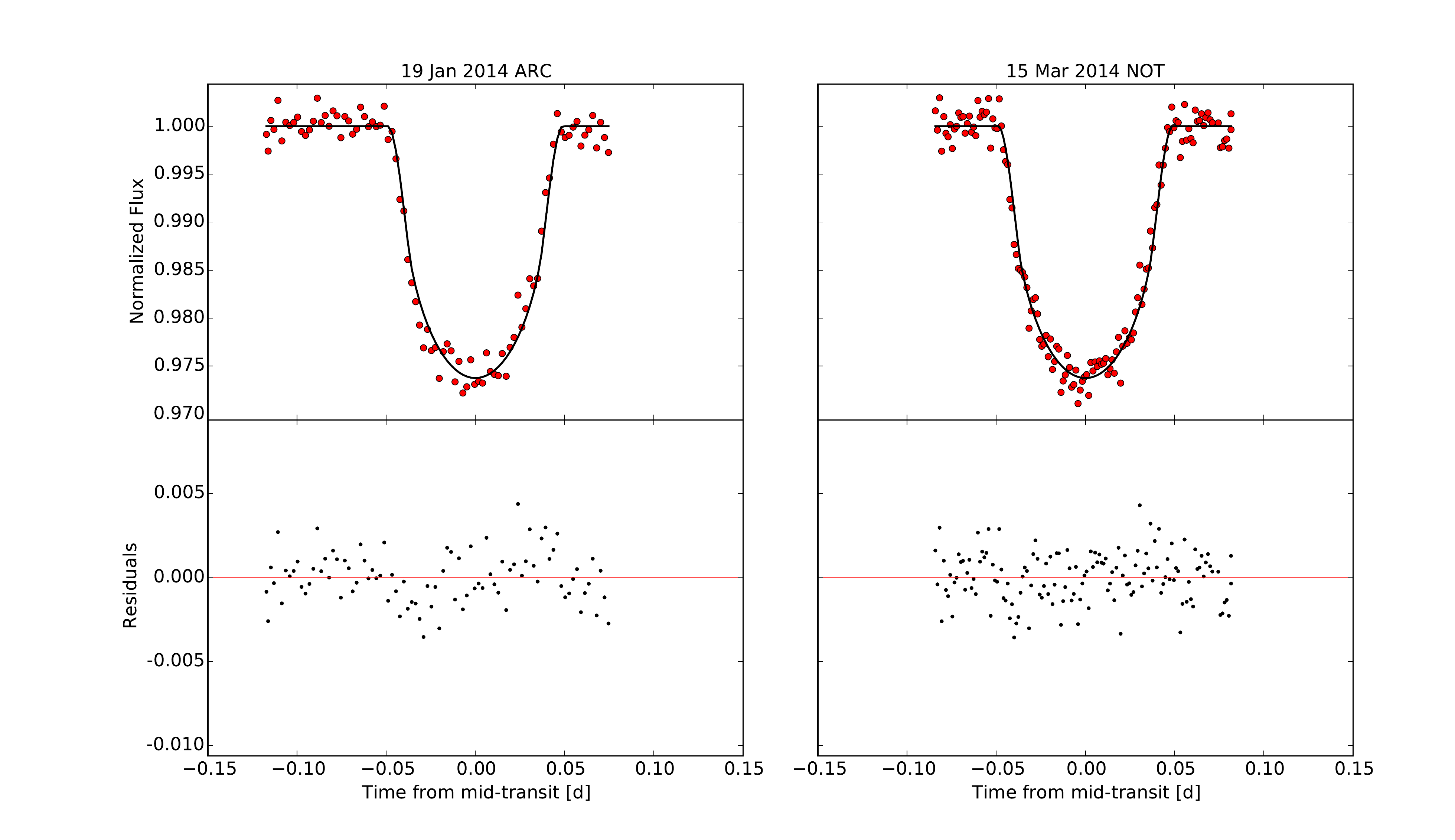}
\caption{Reanalyzed, detrended literature B-band light curves with the associated transit model described in Section \ref{sec_homoreana}. The bottom panel shows the associated residuals. }
\label{b_band_final}
\end{figure*}

\begin{figure*}[h]
\centering
\includegraphics[height=11cm]{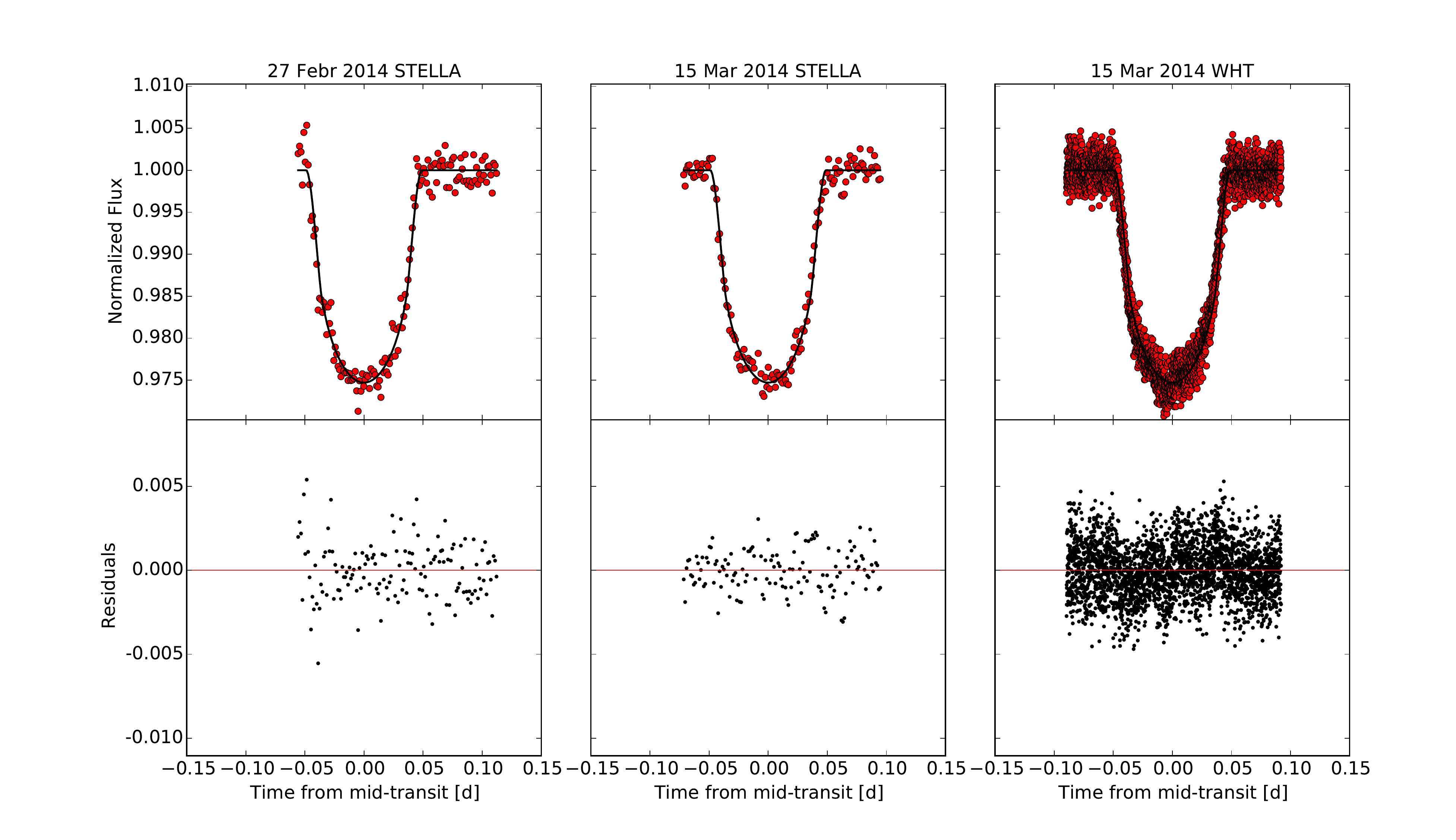}
\caption{Reanalyzed, detrended literature g'-band light curves with the associated transit model described in Section \ref{sec_homoreana}. The bottom panel shows the associated residuals.}
\label{g_band_final}
\end{figure*}

\begin{figure*}[h]
\centering
\includegraphics[height=11cm]{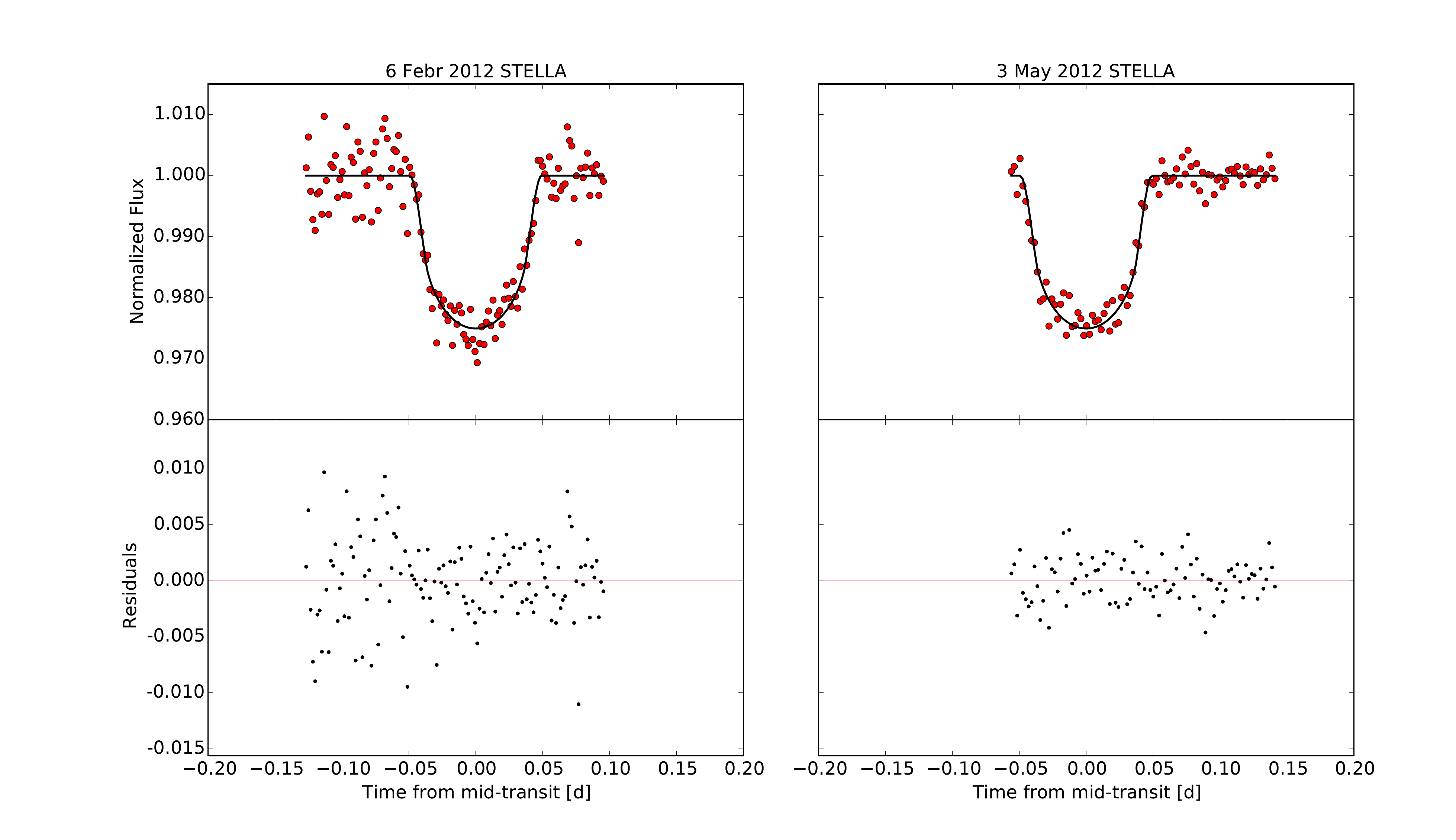}
\includegraphics[height=11cm]{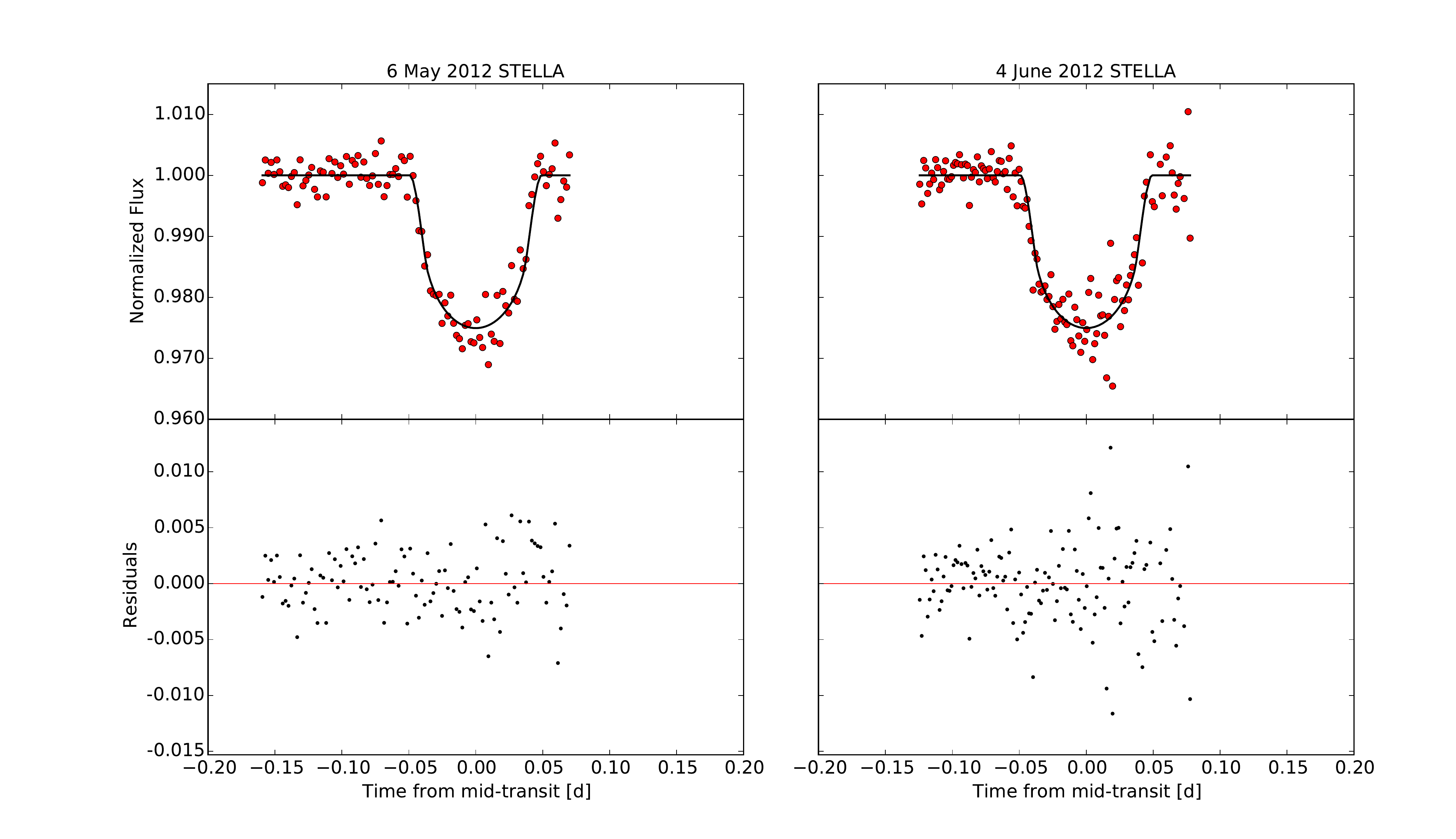}
\caption{Reanalyzed, detrended literature V-band light curves with the associated transit model described in Section \ref{sec_homoreana}. The bottom panel shows the associated residuals.}
\label{V_band_final}
\end{figure*}

\begin{figure*}[h]
\centering
\includegraphics[height=11cm]{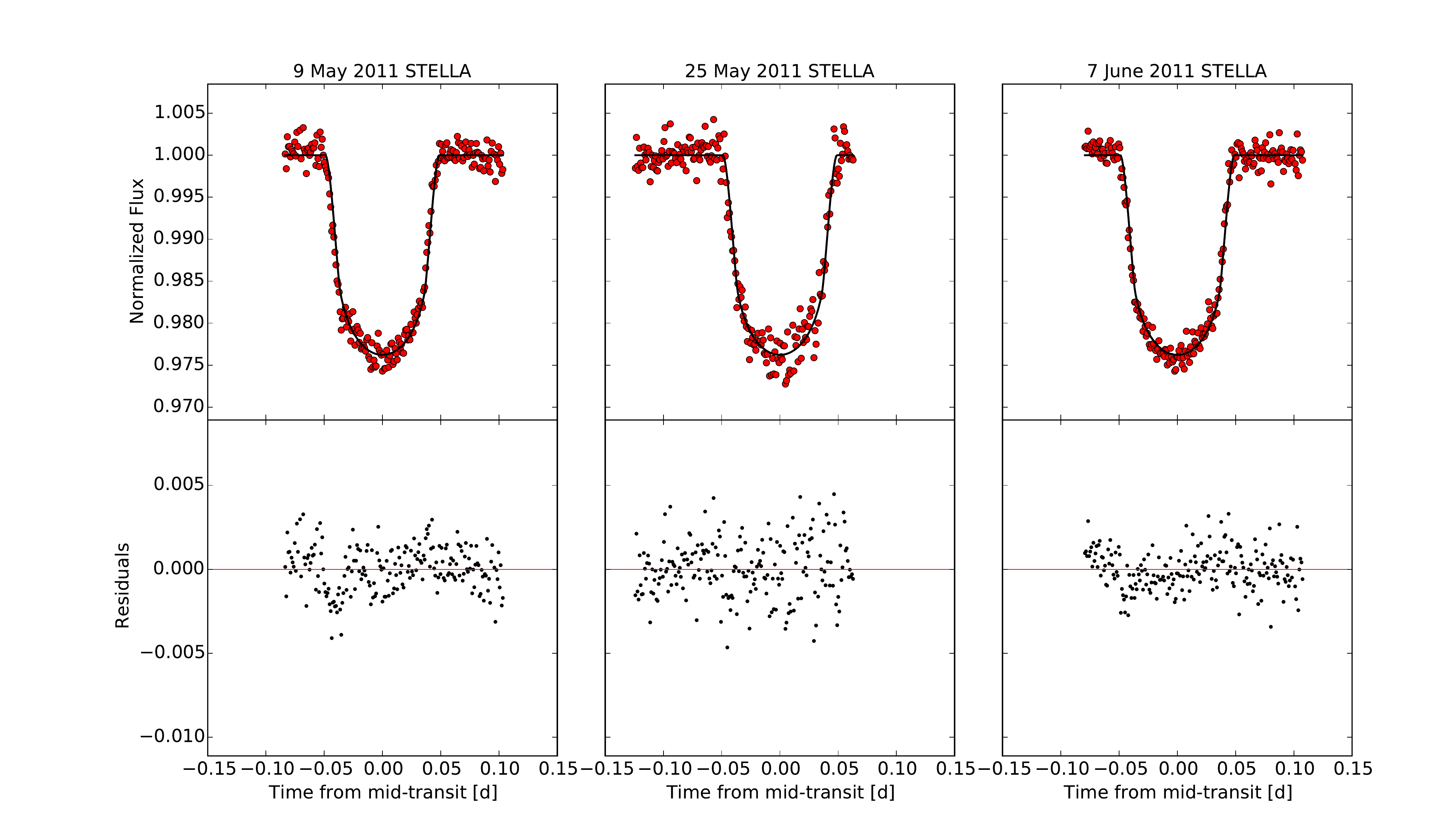}
\includegraphics[height=11cm]{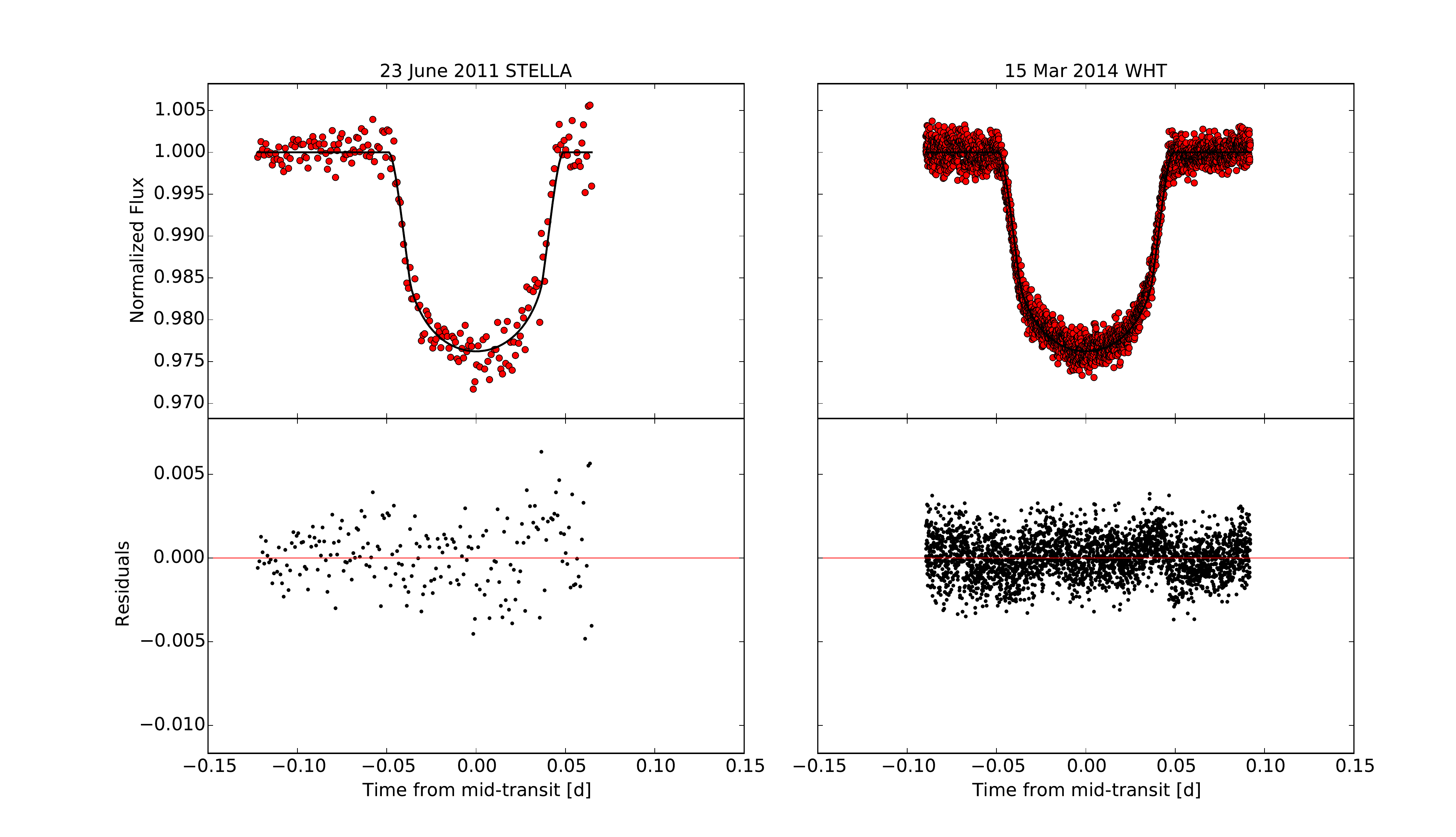}
\caption{Reanalyzed, detrended literature r'-band light curves with the associated transit model described in Section \ref{sec_homoreana}. The bottom panel shows the associated residuals.}
\label{r_band_final}
\end{figure*}

\begin{figure*}[h]
\centering
\includegraphics[height=11cm]{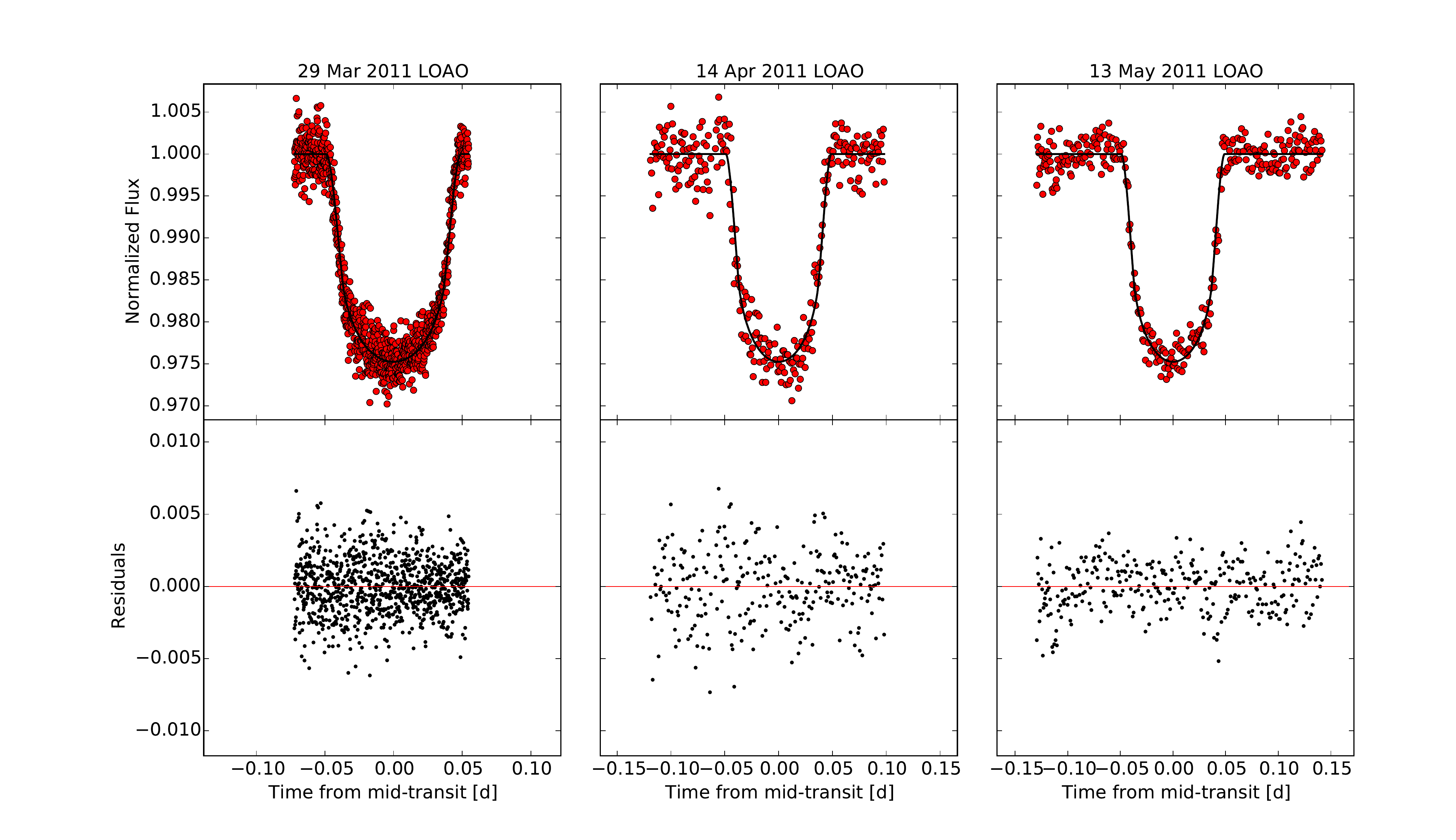}
\includegraphics[height=11cm]{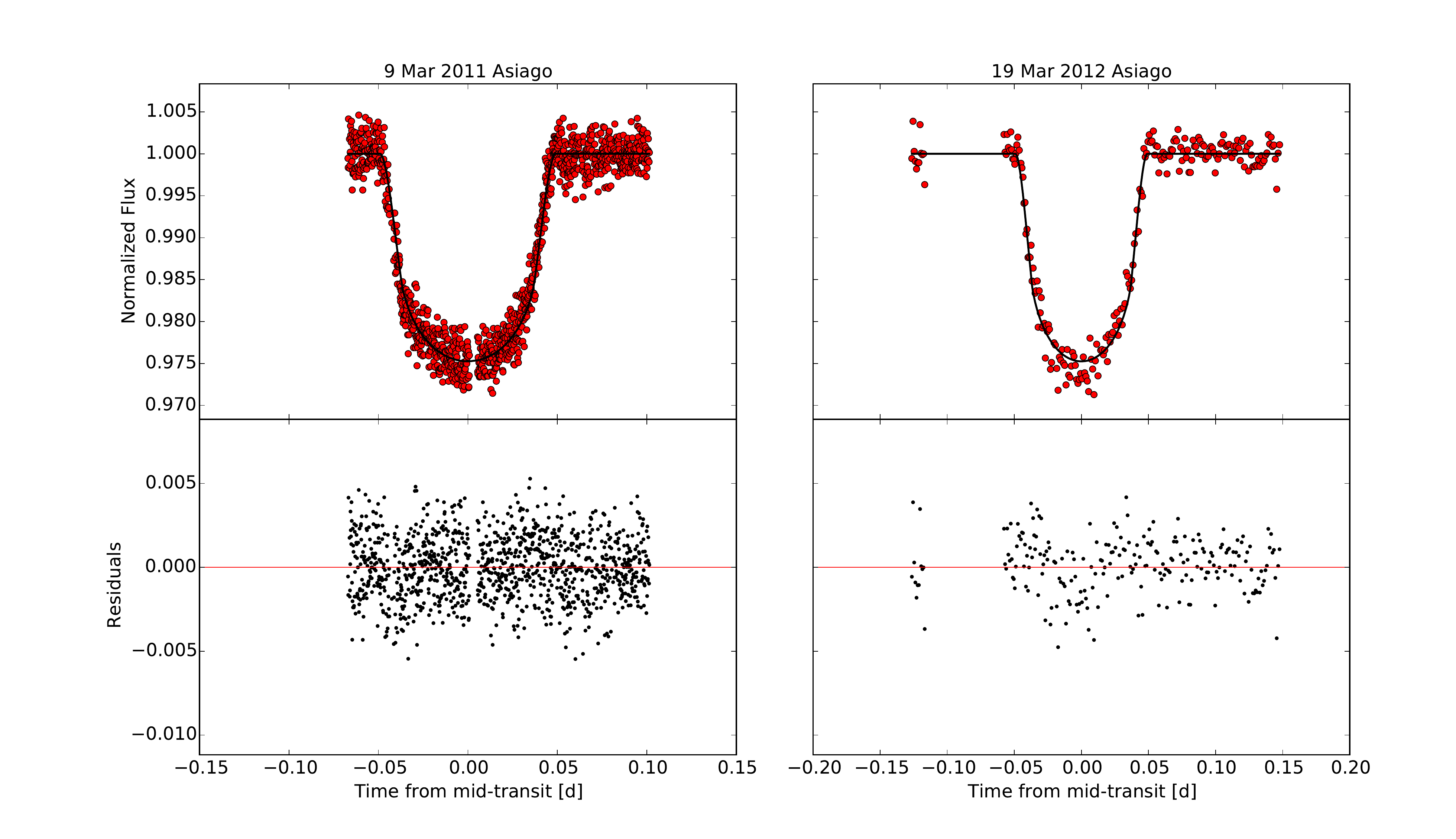}
\caption{Reanalyzed, detrended literature R-band light curves with the associated transit model described in Section \ref{sec_homoreana}. The bottom panel shows the associated residuals.}
\label{R_band_final}
\end{figure*}

\begin{figure*}[h]
\centering
\includegraphics[height=11cm]{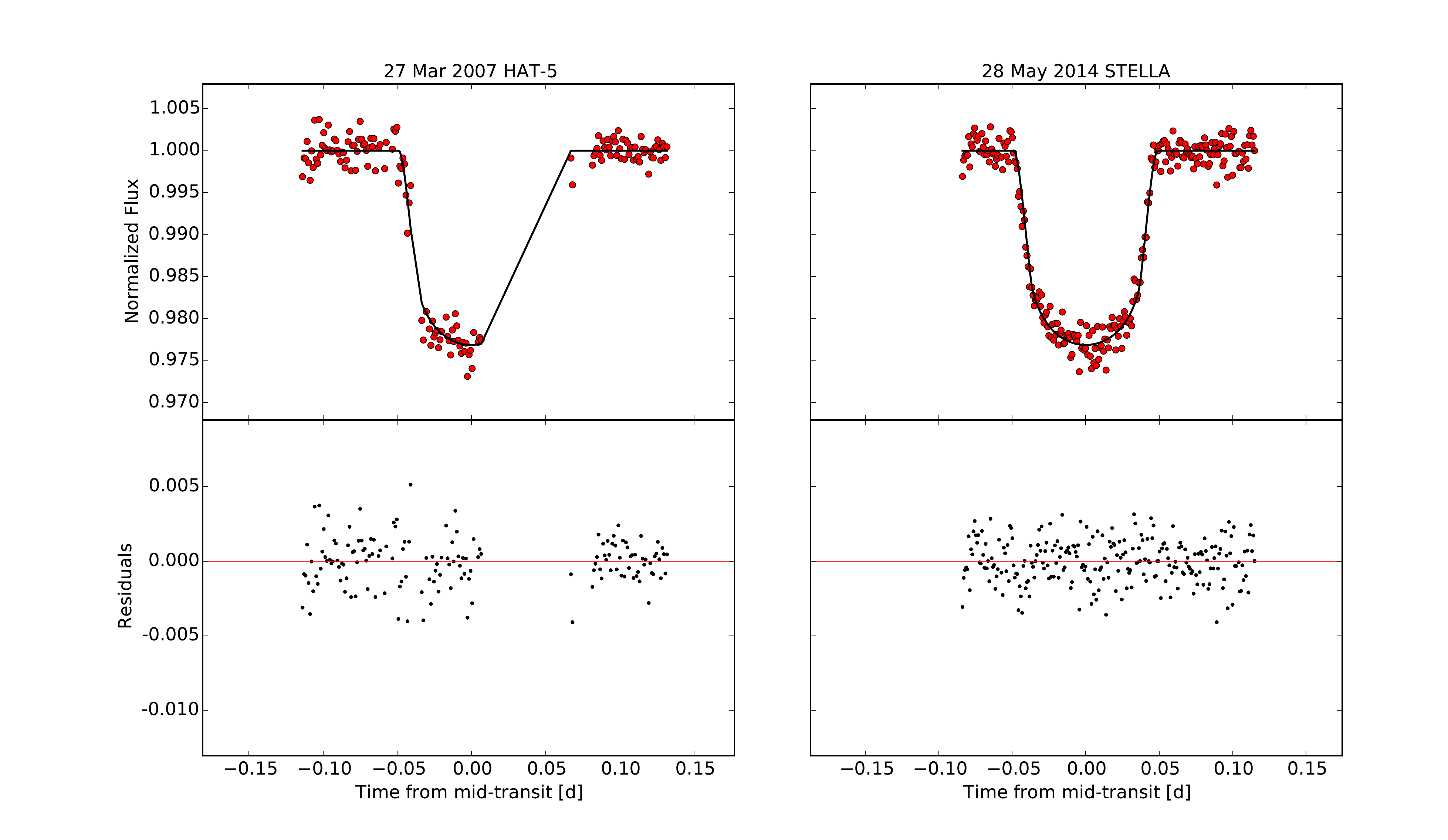}
\includegraphics[height=11cm]{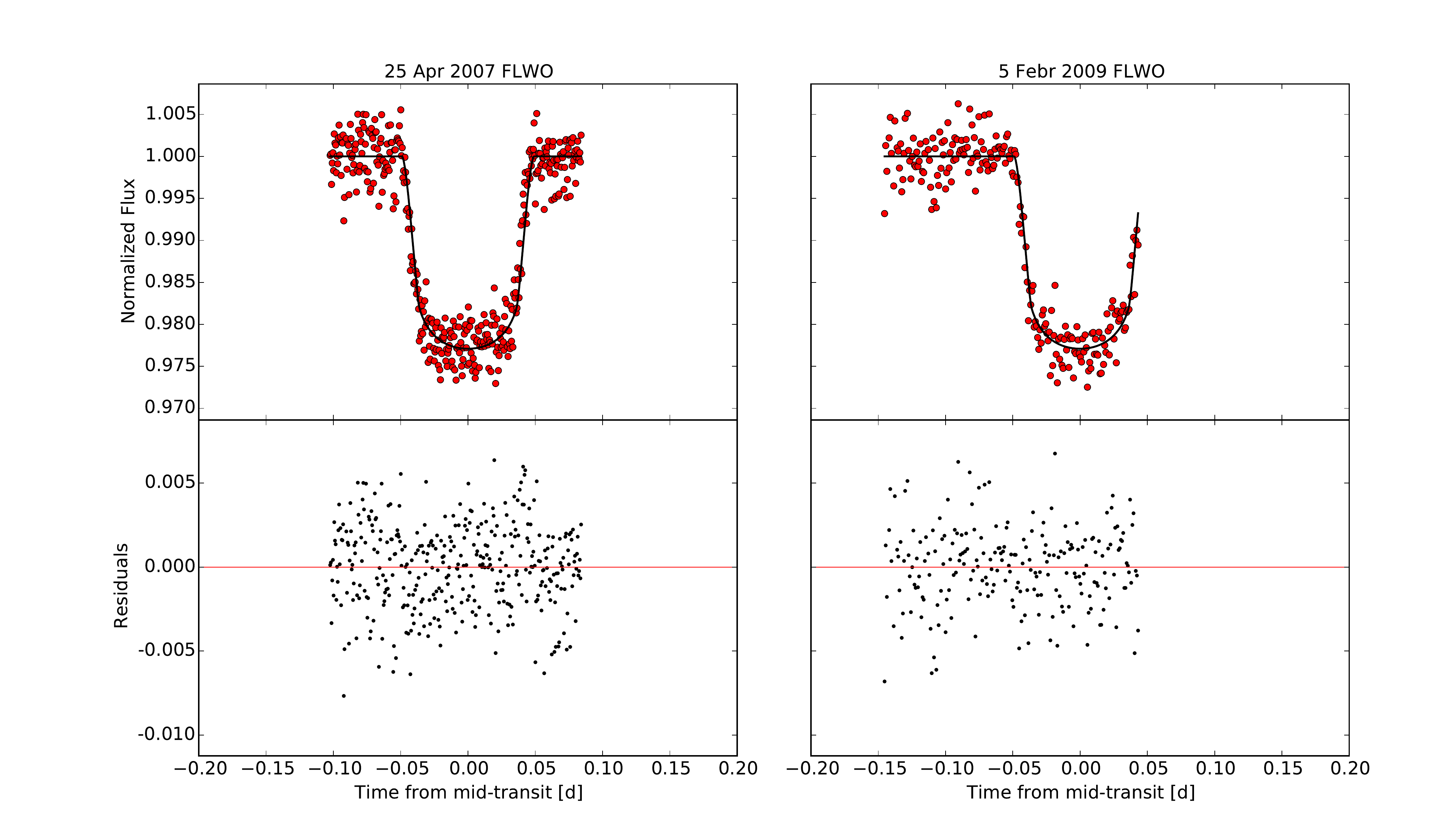}
\caption{Reanalyzed, detrended literature two i'-band (upper part) and two z'-band (lower part) light curves with the associated transit model described in Section \ref{sec_homoreana}. The bottom panels show the associated residuals.}
\label{iz_band_final}
\end{figure*}

%
%

\end{appendix}

\end{document}